\documentclass[aps,
prd,
preprint,
groupedaddress,showpacs,showkeys,12pt]{revtex4-1}
\usepackage{graphicx}
\usepackage{amsmath,amssymb,amsfonts} 
\usepackage{amscd}
\usepackage{amsthm}

\def\proof{{\bf Proof}.\ }

\newtheorem{Lemma}{Lemma}\newtheorem{Theorem}{Theorem}

\newtheorem{Definition}{Definition}
\newtheorem{Remark}{Remark}
\begin{document}
\title{\Large On Concept of Mechanical System}
\author{Al Cheremensky}%
\affiliation{IM--BAS, 4 Acad. G. Bonchev Str., Sofia--1113, Bulgaria\\
E--mail: {cheremensky@yahoo.com}}

\begin{abstract}

\noindent
The paper 
gives a  screw axiomatics of rational mechanics, namely: 
\begin{enumerate}{\em \parskip -0.025cm \vspace{-2pt}
	\item introduces the main measures of mechanics: the mass measure, the scalar and (vector) screw  measures of motion, the (vector) screw measure of impressed action,  the {increment velocity} of the vector measure of motion,  the (vector) screw measure of {constraint action};
\item postulates the  (stronger) local integral form of conservation law for the vector  measuare of motion (fundamental principle of dynamics)
, and 
	\item defines the central concept of rational mechanics -- {\em mechanical system}  being realized in the form of all classical  mechanical systems (mass--points, rigid bodies, continua, point--bodies, {\em etc.}). }
\end{enumerate}

The presentation is based on new notions of vector calculus -- homogeneous and inhomogeneous vector and tensor slider--functions and screw measures.

\end{abstract}
 
\maketitle

%
\medskip

{\large \bf 1. Elements of rational mechanics}

Due to \cite{Truesdell} `{\em Rational Mechanics is the part of mathematics that provides and develops logical models for the enforced changes of place and shape we see everyday things suffer.  $\ldots$Mechanics does not study natural things directly. Instead, it considers bodies, which are mathematical concepts designed to abstract some common features of many natural things. One such feature is the mass assigned to each body. Always, a natural body is at any one instant found to occupy a set of places; that set is the shape of that body at that instant$\ldots$ The change of shape undergone by a body from one instant to another is called the motion of that body. $\ldots$motions of bodies are conceived as resulting from or at least being invariably accompanied by the action of forces$\ldots$ Mechanics relates the motions of bodies to the masses assigned to them and the forces that act on them. Bodies are encountered only in their shapes. Masses and forces, therefore, can be correlated with experience in nature only when they are assigned to the shapes of bodies'}
.

It is easy to see that the statements quoted above lead to the contradiction as the working tools of mechanics are connected with the measures carried over to body shapes, instead of bodies. This contradiction is assumed to be removed by means of carrying over the forces acting on bodies to those shapes `in some special way' \cite{Truesdell}.

Furthermore \cite{Truesdell}:  {\em $\ldots$ the basic categories of mechanics have dual character: on the one hand, in essence such concepts as `force' and `mass' are abstract mathematical categories. On the other hand, they appeal to natural objects with which human practice deals: `force' represents a measure of object interactions, measured by any physical means; `mass' -- quantity of the substance containing in an object'.}

In this connection, it is of importance to notice that, due to this dual character, `body' is considered as a natural thing or as the mathematical concept `designed to abstract some common features of many natural things', but not only -- mass.

It is the concept of body ({\em mechanical system}) that  we shall refine as a mathematical category in this paper.

Below,  we shall use the set ${\bf R}$ of all real numbers and 
$n$--dimensional affine space ${\bf A}_n$ modeled on $n$--dimensional vector space ${\bf V}_n$. 
\newpage
{\bf 1.1.  Slider--functions and screw measures}

We shall introduce the notions of vector and tensor sliders being a key moment  for the theory of mechanical systems which is represented below.

Due to the Great Soviet Encyclopedia, v. 5 (Moscow: Soviet Encyclopedia, 1971), `screw calculus is the section of vector calculus in which operations over screws are studied. Here the screw is called the pair of vectors $\{\vec p, \vec q\}$, bounded at a point $O$ and satisfying to conditions: at transition to a new point $O{\hskip .02cm}'$ the vector $\vec p$ does not change, and the vector $\vec q$ is replaced with a vector $\vec {q}{\hskip .075cm}' =\vec q-\overrightarrow{OO{\hskip .02cm}'}\times \vec p$ where $\times$ means cross--product. The notion of the screw is used in the mechanics (the resultant $\vec f$ of a force system $\{\vec {f}_i\}$ and its main moment $\vec m$ form the screw $\{\vec f, \vec m\}$), and also in geometry (in the theory of ruled surfaces)'.

Note that we do not support the idea to use the name `torser' from the French `torseur' instead of `screw' \cite{Berthelot}.
 
Let us add more details to this notion.
\medskip 

{\bf 1.1.1.  Slider vector--functions}.  
Assume that there are vectors $\vec p_x $ and $\vec {q}_x$ bounded at a given point $x\in {\bf A}_3$, and at any point $y\in {\bf A}_3$
\begin{equation}
\vec p_y=\vec p_x, \quad \vec q_y=\vec {q }_x+r_{yx}^\times \vec p_x
\label{ 1}\end{equation}
where 
 $r_{yx}^\times$ is the spin--tensor generated by the vector $\vec r_{yx}=\overrightarrow{yx}$.

\begin{Definition} The field $l ^{\hskip .02cm p_x ,q_x}=\{\vec p_y, \vec q_y, \forall y\in {\bf A}_3\}$ is called {\em slider vector--function} or, briefly, {\em slider}  while $l_y ^{\hskip .02cm p_x ,q_x}\stackrel{{def}}{=} \{\vec p_y, \vec q_y\}$ is known as {\em reduction} of the slider w.r.t. the {\em reduction point} $y\in {\bf A}_3$. \end{Definition}
 
 A slider is called {\em homogeneous} if $ \vec q_x =0$. In this case we shall use the notation $l^{\hskip .02cm p_x}$.
 
If one marks coordinate columns of vectors in a Cartesian frame $\mathcal E_0=(O_0,{\bf e}_{0})$ ($O_0$ is its origin  and ${\bf e}_{0}$ is its base) with the superscript ${}^0$, then  $l^{\hskip .02cm p_x ,q_x,0}{=}\{{p }^{\hskip .02cm 0}_y , q^{\hskip .02cm 0}_y, \forall y\in {\bf A}_3\}$ is the coordinate representation of the slider  $l^{\hskip .02cm p_x ,q_x}$. In order to apply the matrix tools one may use the following coordinate columns $l_y ^{\hskip .02cm p_x ,q_x, wr,0}={\rm col}\{ {p }^{\hskip .02cm 0}_y , q^{\hskip .02cm 0}_y
\}$ and $l_y ^{\hskip .02cm p_x ,q_x, tw,0}={\rm col}\{ q^{\hskip .02cm 0}_y, {p }^{\hskip .02cm 0}_y
\}$ known as {\em wrench} and {\em twist} (in the space ${\bf E}_6$), respectively.
\medskip

{\bf 1.1.2. Screw measures}. 
Let $\sigma _3$ be $\sigma $--algebra of subsets in ${\bf A}_3$. Introduce the following Borel measure 
\[
\mu (A)=\mu _{ac}(A)+\mu _{pp}(A), \quad A\in \sigma _3 
\]
where $\mu _{ac}(A)$ is the absolutely continuous component w.r.t. Lebesgue
measure $\mu _3$ and $\mu _{pp}(A)$ is the pure point (discrete) component presented as $\mu _{pp}(A)=\sum_k\mu _ k$ for points ${x }_k\in A$ whose are called {\em pure}, the others being called {\em continuous} \cite{Reed}. 
\begin{Definition}
	Let $\chi _{\hspace {-.05cm}_{A}}$ be the {\em characteristic function} of $A$. The Lebesgue--Stieltjes integral
	\begin{equation}
		\pi(A)=\int \hspace{-.05cm} \chi _{\hspace {-.05cm}_{A}}l^{p_x,q_x}\mu (dx), \quad A\in \sigma _3\label{screw}\vspace{-3pt}
\end{equation}
		is called {\em signed screw measure}  or {\em screw}.  Signed measure is the generalization of the notion of measure by allowing it to have negative values \cite{Evans}.
\end{Definition}
We shall use this name for surface Lebesgue--Stieltjes integrals, too.

The screw measure is a screw in the sense of the Encyclopedia definition (if $\{\vec p, \vec q\}\stackrel{{def}}{=}\pi_0(A)$ at the point $O$ then there is $\pi_0{{\hskip -.15cm}'}(A)\stackrel{{def}}{=}\{\vec p, \vec {q}{\hskip .05cm}'\}$ at the point $O{\hskip .02cm}'$ where $\vec {q}{\hskip .05cm}'=\vec q-\overrightarrow{OO{\hskip .02cm}'}\times \vec p\hskip .05cm$). 

Screws generated by homogeneous (inhomogeneous) sliders will be called homogeneous (inhomogeneous).

Define the triple of orthogonal unit vectors 
$\vec e_1$, 
$\vec e_2$ and 
$\vec e_3$ in the $3$--dimensional space ${\bf V}_3$. Let us introduce $6$ screws such that at a point $y\in {\bf A}_3$ their reductions are defined as follows\vspace{4pt}
\[
{\mathfrak e}_1=\begin{pmatrix} \vec e_1 \vspace{-4pt}\\ \vec o
\end{pmatrix},\ 
{\mathfrak e}_2=\begin{pmatrix} \vec e_2 \vspace{-4pt}\\ \vec o
\end{pmatrix},\ 
{\mathfrak e}_3=\begin{pmatrix} \vec e_3 \vspace{-4pt}\\ \vec o
\end{pmatrix},\ 
{\mathfrak e}_4=\begin{pmatrix} \vec o \vspace{-4pt}\\ \vec e_1
\end{pmatrix},\ 
{\mathfrak e}_5=\begin{pmatrix} \vec o \vspace{-4pt}\\ \vec e_2
\end{pmatrix},\ 
{\mathfrak e}_6=\begin{pmatrix} \vec o \vspace{-4pt}\\ \vec e_3
\end{pmatrix}\vspace{5pt}
\]
 where $\vec o\in {\bf V}_3$ is the null vector.
 
 As any screw is defined in the unique way by its reduction at some point, these $6$ screws generate a basis of $6$--dimensional vector space 
 (see also \cite{Berthelot}).
\medskip

{\bf 1.1.3. Multiplicative groups  of motions}.  
In ${\bf A}_3$ let us introduce the Cartesian frames ${\cal E}_p={\cal E}(O_p,{\bf e}_ p)$ and ${\cal E}_k={\cal E}(O_k,{\bf e}_ k)$ with the origins $O_p$ and $O_k$ and the bases ${\bf e}_{p}$ and ${\bf e}_{ k}$ 
where $ p$ and $ k$ are naturals.

Define the rotation matrices $C_{0,p}$ and $C_{p,k}$ defining orientations of the Cartesian frames ${\cal E}_p$ and ${\cal E}_k$ w.r.t ${\cal E}_0$ and ${\cal E}_p$, respectively. Then $C_{0, p}C_{p,k}= C_{0,k}$ and for any free vector $\vec {\lambda}$ there are the following relations
\[\lambda^0= C_{0,p}\lambda^p, \quad \lambda^p= C_{p,k}\lambda^k 
\] 
where $\lambda^0$, $\lambda^p $ and $\lambda^ k \in {\bf E}_3$ are the coordinate columns of the vector $\vec {\lambda}$ in the bases ${\bf e}_{0}$, ${\bf e}_ p$ and $
{\bf e}_{ k}$. Hence we have also $\lambda^0= C_{0,k}\lambda^ k$.

Introduce the radius--vectors $\vec {r}_x$ and $\vec {r}_{p,x}\in {\mathbf V}_3$ of
a point $x\in {\bf A}_3$ w.r.t. the origins $O_0$ and $O_p$, respectively. Define $\vec d_{0,p}=\vec {r}_x-\vec {r}_{p,x}$. Then we may 
represent the relation $\vec {r}_x=\vec {d}_{0,p}+\vec {r}_{p,x}\in {\bf V}_3$ in the  frame ${\cal E}_0$ as $r^0_x=d^0_{0,p}+C_{0,p}r_{p,x}^p$. Let the point $x$ be  immovable in ${\cal E}_p$ then $r_{p,x}^p$ is time--constant.

With differentiating the last relation we have 
$v^0_x=v_{0,p}^{0}+C_{0,p}^{\hspace{.02cm }{}^\centerdot}r_{p,x}^p$ where $\vec v_x=\vec r_x^{\hspace{.1cm }{}^\centerdot }$ is the velocity of $x$  and $\vec v_{0,p}=\vec d_{0,p}^{\hspace{.1cm }{}^\centerdot }$ is {\em translation velocity} of ${\cal E}_p$ w.r.t. $O_0$, respectively. Here to honor Newton, we use the superscript ${}^\centerdot$ for full derivatives by $t$, {\em e.g.}, for any function $A=A(x(t),t)$ we have $A^ {\hspace{.02cm }{}^\centerdot}=\frac{\partial}{\partial t}A+({\rm div}\hspace {.02cm} A)\hspace {.05cm} x^ {\hspace {.02cm} \centerdot}(t)$.

Hence 
\begin{equation}
	v^p_x=v_{0,p}^{p}+C_{p, 0}C_{0,p}^{\hspace{.02cm }{}^\centerdot}r_{p,x}^p
	\label{cross}
\end{equation}
For any vector $f=\begin{pmatrix}
f _1 \vspace{-3pt}\\ 
f _2 \vspace{-3pt}\\ 
f _3 
\end{pmatrix}\in {\bf E}_3$ introduce the cross product matrix\vspace{-15pt} \begin{equation}
f^{\times}\stackrel{\mathrm{def}}{=} \hspace{-0.12cm}\begin{bmatrix}
0 & -f_3 & f _2 \\ 
f_3 & 0 & -f _1 \\ 
-f_2 & f _1 & 0
\end{bmatrix}\label{ 2g}
\end{equation}
Thus we may define (in ${\cal E}_p$) \cite{Lurie}:
\vspace{-17pt}\begin{quote}{\parskip -.24cm
\item  \hskip -.8cm  --- \thinspace \thinspace  \thinspace \thinspace 
the coordinate column $d _{0,p}^p\in {\bf E}_3$ of the {\em translation} vector of ${\cal E}_p$ w.r.t. ${\cal E}_{0}$ in ${\cal E}_p$;
\item  \hskip -.8cm  --- \thinspace \thinspace  \thinspace \thinspace 
the coordinate column $v _{0,p}^p\in {\bf E}_3$ being known as {\em quasi--velocity} of the translation of ${\cal E}_p$ w.r.t. ${\cal E}_{0}$ in ${\cal E}_p$;
\item  \hskip -.8cm  --- \thinspace \thinspace  \thinspace \thinspace 
the cross product matrix $\omega _{0,p}^{p\times} \stackrel{\rm { def}}{=} C_{p, 0}C_{0,p}^{\hspace{.02cm }{}^\centerdot}$ where the triple $\omega _{0,p}^p\in {\bf E}_3$ is known as {\em quasi--velocity} of rotation of ${\cal E}_p$ w.r.t. ${\cal E}_{0}$ (calculated in ${\cal E}_p$).}\vspace{-3pt}
\end{quote}
The algebraic quantity $\omega _{0,p}^p$ corresponds to the geometrical one, {\em i.e.},  the instantaneous angular 
velocity $ \vec \omega _{0,p}\in {\bf V}_3$. 
It defines the rotation axis of ${\cal E}_p$. 

Introduce following matrices
\begin{equation}
{C}_{0,p}^\otimes 
= 
\begin{bmatrix}
C_{0,p} &\ O \\ 
O &\ C_{0,p}
\end{bmatrix}
, \quad 
{ D}^0_{0,p}=
\begin{bmatrix}
I &\ O\\ 
 \hspace{ .15cm}d_{0,p}^{0\times } &\ \hspace{ .15cm}I
\end{bmatrix}, \quad { D}^p_{0,p}=
\begin{bmatrix}
I &\ O\\ 
 \hspace{ .15cm}d_{0,p}^{p\times } &\ \hspace{ .15cm}I
\end{bmatrix}
\label{ 3g}
\end{equation}
\begin{Theorem} {\rm \cite{Konoplev1999}} A given screw $\pi$, $\pi_0 ^{wr, 0}= {L}^{wr}_{0,p}\pi_p^{wr ,p}$  where
$\pi_0 ^{wr, 0}$ and $\pi_p ^{wr ,p}$ are wrenches of $\pi$ computed in the frames  ${\cal E}_0$ and ${\cal E}_p$, respectively,  the matrix ${ L}^{wr}_{0,p}$ has the representation 
\begin{equation}
	{ L}^{wr}_{0,p}={ D}^0_{0,p}{C}_{0,p}^\otimes ={C}_{0,p}^\otimes { D}^p_{0,p}\label{66}
\end{equation}
 and belongs to the multiplicative group ${\mathcal L}^{wr}({\mathcal R}, 6)$ such that 
\begin{equation} 
{ L}^{{wr}\centerdot }_{0,p}={ L}^{wr}_{0,p}\varPhi_{0,p}^{wr}=\varPsi_{0,p}^{wr}{ L}^{wr}_{0,p}, \quad \varPhi_{0,p}^{wr}=
\begin{bmatrix}
 \omega_{0,p}^{p\times} & O \\ 
 v_{0,p}^{p\times} & \omega_{0,p}^{p\times}
\end{bmatrix}, \quad \varPsi_{0,p}^{wr}=
\begin{bmatrix}
 \omega_{0,p}^{0\times} & O \\ 
 v_{0,p}^{0\times} & \omega_{0,p}^{0\times}
\end{bmatrix}
\label{ 41}
\end{equation}
\end{Theorem}
\proof Representation (\ref{66}) follows directly from the screw definition.

Consider the case where ${ L}^{{wr}}_{0,p}={C}_{0,p}^\otimes { D}^p_{0,p}$. Then  from (\ref{ 3g}) follows that 
 ${ L}^{{wr}\centerdot}_{0,p}={C}_{0,p}^{\otimes \centerdot} { D}^p_{0,p}+{C}_{0,p}^\otimes { D}^{p\centerdot}_{0,p}=({C}_{0,p}^{\otimes \centerdot} { D}^p_{0,p}{ D}^p_{p,0}{C}_{p,0}^\otimes +{C}_{0,p}^\otimes { D}^{p\centerdot}_{0,p}{ D}^p_{p,0}{C}_{p,0}^\otimes ){C}_{0,p}^\otimes { D}^p_{0,p}
=({C}_{0,p}^{\otimes \centerdot}{C}_{0,p}^\otimes +{ D}^{0\centerdot}_{0,p}){C}_{0,p}^\otimes { D}^p_{0,p}=\varPsi_{0,p}^{wr}{ L}^{wr}_{0,p}
$.

The matrices of the kind ${ L}^{wr}_{0,p}={C}_{0,p}^\otimes { D}^p_{0,p}$ 
form a group because there are $L^{{wr}}_{0,p}L^{{wr}}_{p,k}={C}_{0,p}^\otimes {C}_{p,k}^\otimes {C}_{k,p}^\otimes { D}^p_{0,p}{C}_{p,k}^\otimes { D}^k_{p,k}= 
{C}_{0,k}^\otimes { D}^k_{0,p}{ D}^k_{p,k}={C}_{0,k}^\otimes { D}^k_{0,k}=L^{{wr}}_{0,k}$  and $%
L_{0,p}^{{{wr}}, -1}=({C}_{0,p}^\otimes { D}^p_{0,p})^{-1}=(T^p_{0,p})^{-1}C_{0,p}^{\otimes, T}=T^p_{p,0}C_{p,0}^{\otimes}
=C_{p,0}^{\otimes}T^0_{p,0}=L^{{wr}}_{p,0}$.

Consider the case where ${ L}^{{wr}}_{0,p}={ D}^0_{0,p}{C}_{0,p}^\otimes$. Then  from (\ref{ 3g}) follows that 
${ L}^{{wr}\centerdot}_{0,p}={ D}^{0\centerdot}_{0,p}C_{0,p}^\otimes +{ D}^0_{0,p}{C}^{\otimes \centerdot}_{0,p}={ D}^0_{0,p}
{C}_{0,p}^\otimes ({ C}_{p,0}^\otimes { C}^{\otimes \centerdot}_{0,p}$ $+{ C}_{p,0}^\otimes { D}^0_{p,0}{ D}^{0\centerdot
}_{0,p}{ C}_{0,p}^\otimes ) ={ L}^{wr}_{0,p}\varPhi_{0,p}^{wr}$.

The matrices of the kind ${ L}^{wr}_{0, p}={ D}^0_{0,p}{C}_{0,p}^\otimes$ 
form a group because there are $L^{{wr}}_{0,p}L^{{wr}}_{p,k}= T_{0,p}^0C_{0,p}^\otimes T_{p,k}^p C_{p,k}^\otimes=  T_{0,p}^0T_{p,k}^0C_{0,k}^\otimes ={ D}^0_{0,k}C_{0,k}^\otimes=L^{{wr}}_{0,k}$ and $%
L_{0, p}^{{{wr}}, -1}=(T^0_{0, p}C_{0, p}^{\otimes})^{-1}=C_{p,0}^{\otimes}(T^0_{0, p})^{-1}=C_{p, 0}^\otimes T^0_{p, 0}C_{0, p}^{\otimes, T}C_{0,p}^\otimes =T^p_{p, 0}C_{p, 0}^\otimes=L^{{wr}}_{p, 0}$. 

As result we have the following relations\vspace{-3pt}
\begin{equation}
{ L}^{{wr}\centerdot}_{0,p}={ L}^{wr}_{0,p}\varPhi_{0,p}^{wr},\quad  { L}^{{wr}}_{0,p}={C}_{0,p}^\otimes { D}^p_{0,p}\label{p}\vspace{-3pt}
\end{equation}
being matrix--functions of $d_{0,p}^{p\times }$, $v_{0,p}^{p\times}$, and $\omega_{0,p}^{p\times}$.

The similar statement $\pi_p ^{tw, p}= {L}^{tw}_{0,p}\pi_0^{tw ,0}$ is true for twists where we have the matrix 
\begin{equation}
	{ L}^{tw}_{0,p}=\begin{bmatrix} O &\hspace {.25cm} I \\ I 
	&\hspace {.25cm} O \end{bmatrix}{ L}^{wr}_{0,p}\begin{bmatrix} O &\hspace {.25cm} I \\ I 
	&\hspace {.25cm} O \end{bmatrix}\label{L^{tw}}
\end{equation} belongs to the multiplicative group ${\mathcal L}^{tw}({\mathcal R}, 6)$ such that $
{ L}^{{tw}\centerdot }_{0,p}=\varPsi_{0,p}^{tw}{ L}^{tw}_{0,p}$, 
$ \varPsi_{0,p}^{tw}=- \varPsi_{0,p}^{wr, T}
$ and $
{ L}^{{tw}\centerdot }_{0,p}=\varPhi_{0,p}^{tw}{ L}^{tw}_{0,p}$, 
$ \varPhi_{0,p}^{tw}=- \varPhi_{0,p}^{wr, T}
$. 

Note that in contrast to the groups of motions in the $3$--dimensional space the groups ${\mathcal L}^{wr}({\mathcal R}, 6)$ and ${\mathcal L}^{tw}({\mathcal R}, 6)$ are multiplicative. \medskip

{\bf 1.1.4. Slider tensor--functions}. 
The slider notion is based on the pair of vector--functions $\vec p_x$ and $\vec {q }_x$. That is why these sliders are called {\em vector} ones. If we replace these vector--functions with tensors ${\mathcal P}_x$ and ${\mathcal Q}_x$ of II rank, then the corresponding sliders will be called {\em tensor} ones. \newpage

{\bf 1.2.  Main concepts and structures of mechanics}

 In what follows, we shall use {\em Galilean space--time} \cite{Arnold} introduced as the quadruple ${\bf G} = \{{\bf V}_4, {\bf A}_4, \tau, g\}$ where \vspace{-7pt}
\begin{enumerate}{\parskip -.05cm \em 
\item $\tau \hspace{-.15cm}: {\bf V}_4 \rightarrow {\bf V}_1$ is a surjective linear mapping called {\em time} one, and\vspace{-7pt}
\item $g=\langle \cdot,\cdot\rangle $ is an inner product on ${\rm ker} \{\tau \}$ $(={\bf V}_3)$.} \vspace{-7pt}
\end{enumerate}
The points of ${\bf A}_4$ are called {\em world points} or {\em events}. The number $ \tau (b - a) $ is called {\em time interval} between events $a $ and $b\in {\bf A} _4$. These events $a $ and $b\in {\bf A} _4$ are called {\em simultaneous} if $ \tau (a-b) =0$. The set of simultaneous events forms $3$--dimensional affine space $ {\bf A} _3$ in affine space $ {\bf A} _4$.

The inner product $\langle \cdot,\cdot\rangle $ (in Galilean space--time) enables one to pass from the space ${\bf V}_3$ to {\em Euclidean} space ${\bf E}_3$ with the norm ${\| \vec x\|} = \sqrt{\langle \vec x,\vec x\rangle }$ and to introduce {\em Cartesian} frame $\mathcal E_0$ in ${\bf A}_3$ (with the origin $O_0\in {\bf A} _3$).

The basis ${\bf e}_0=(e_{0, 1}, e_{0, 2}, e_{0, 3})$, $
e_{0, 1}= {\rm col}\{1, 0, 0\}$, $
e_{0, 2}={\rm col}\{0, 1, 0\}$, $e_{0, 3}={\rm col}\{0, 0, 1\}$ is called {\em canonical}. Free vectors and their coordinate columns in the canonical basis ${\bf e}_0$ are known with the same name {\em vector} as elements of the vector spaces ${\bf V}_3$ and ${\bf E}_3$. In what follows, one uses the number space ${\bf E}_3$ as representation of ${\bf V}_3$. 

Any set ${\bf T}\subset {\bf R}$ may be used for {parameterization} of the image of $\tau$ with $\sigma $--algebra $ \sigma _t$ of subsets in ${\bf R}$. Values of parameter $t\in {\bf T}$ are called {\em instants}. We shall assume that there is defined $\sigma$-algebra $\sigma_t$ and the Lebesgue measure $\mu (dt)$ on the set ${\bf T}$. 

Let us introduce following notions \cite{Truesdell}. 
{\em World--line} is a curve in ${\bf A}_4$ whose image in ${\bf A}_3\times {\bf T}$ associates one point $x(t) \in {\bf A}_3$ to each instant $t \in {\bf T}$. A collection of non--intersectional world--lines forms {\em world--tube}. 
Here `Intersections of world--lines represent collisions or the creation or destruction of bodies or elements of bodies. In specific mechanical theories such intersections are usually excluded (the principle of impenetrability) altogether or allowed as exceptional cases subject to specified conditions' \cite{Truesdell}. 

Henceforth we shall name some  world--tube $\tilde{\bf \Lambda}\subset {\bf A}_4$ as {\em universe}. As in {\em probability theory} \cite{Kolmogorov}, the universe is separately specified for every mechanical problem. 

A given world--tube ${\bf \Lambda}\subset \tilde{\bf \Lambda}$, the world--tube $ {\bf \Lambda}^{e}=\tilde{\bf \Lambda}\setminus {\bf \Lambda}$ is called {\em environment} of ${\bf \Lambda}$ in the universe. 

The universe $\tilde{\bf \Lambda}$ defines the family $\{\tilde{\Lambda}_t\subset {\bf A}_3$, $t \in {\bf T}\}$, for any world--tube ${\bf \Lambda}\subset \tilde{\bf \Lambda}$ we having  the family $\{{\Lambda_t}\subset \tilde{\Lambda}_t$, $t \in {\bf T}\}$.   We shall assume that the Borel measure introduced above is time--invariant on the sets ${\Lambda}_t$ and, if a point $ {x} (t_\ast) $ of any curve $\{{x }(t) \in \tilde{\Lambda}_t, t \in {\bf T}\}$ is pure (or continuous) at some time instant $t_\ast $, all points of this curve are also pure (or continuous). 

Let us introduce kinematical and dynamic structures in the universe.

For each point $x (t) \in \tilde{\Lambda}_t$, $t \in {\bf T} $, the radius--vector $ \vec {r} _ {x} (t) = \overrightarrow {(O_0, x (t))} $ is called {\em position} of the point, and a vector $ \vec {v} _ {x}= \vec {v}(x(t),t) \stackrel {def} =\vec {r} _ {x} ^ {\hspace {.1cm} \centerdot} (t)
$ is called its {\em velocity} w.r.t. $O_0$  at instant $t\in {\bf T}$.

We shall call {\em mass} the scalar measure ${\mathcal M}({\Lambda_t})$, being continuous w.r.t.  $\mu (dx)$.
Due to Radon--Nicodym theorem the measure may be represented as the following Lebesgue--Stieltjes integral
 \begin{equation}
		{\mathcal M}({\Lambda_t})=\int \hspace{-.05cm} \chi _{\hspace {-.05cm}_{{\Lambda_t}}}\rho _x\mu (dx), \quad \Lambda_t\subset \tilde{ \Lambda}_t
		\label{M}
\end{equation}where $\rho _x$ is the mass density.

Let us define the following scalar measure  \vspace{-3pt} \begin{equation}	 {\mathcal K}({\Lambda_t})=\int \hspace{-.05cm} \chi _{\hspace {-.05cm}_{ \Lambda_t}}k_x\hspace{.01cm} \mu(dx)
 \label{K}\end{equation}
with the density $k_x=\frac{1}2{\langle \vec v_x,\rho _x\vec v_x\rangle }$.

Introduce the following vector and  {screw}: \vspace{-3pt}
\[ \vec{p}_x=
\frac{\partial}{\partial \vec{v}_x} k_x=\rho_x\vec{v}_x,\quad {\mathcal P}({\Lambda_t})=\int \hspace{-.05cm} \chi _{\hspace {-.05cm}_{ \Lambda_t}} l^{\hspace{.02cm}p_x} \mu (dx)\vspace{-5pt}\]
Due to \cite{Konoplev1999} we shall use the notion of {\em bi--measure}: a vector--function ${\it \Phi}(\cdot ,\cdot )$, which is defined on $\sigma _3\times
\sigma _3$ and a screw  measure of the kind (\ref{screw}) by each of arguments, is
called {\em screw bi--measure}.

A bi--measure ${\it \Phi} (A, B)$ is called {
skew} if 
${\it \Phi}(A,B)=-{\it \Phi} (B,A)$ for any $A$ and $B\in \sigma _3$.

Let a skew  screw bi-measure  ${\it \Phi}({\Lambda_t},\Lambda_t^e)$ be homogeneous.  By definition it is the screw w.r.t. every argument. That is why there exists the slider   $l^{\hskip .02cmf_x}$ such that the bi-measure coincides (by the first argument) with the following screw 
 \begin{equation}
{\mathcal F}({\Lambda_t})=\int \hspace{-.05cm} \chi _{\hspace {-.05cm}_{\Lambda_t}} l^{\hskip .02cmf_x}\mu (dx)\stackrel{{def}}{=} {\it \Phi}({\Lambda_t},\Lambda_t^e)
	\label{F}
\end{equation}
and ${\it \Phi}({\Lambda_t^e},\Lambda_t){=}-{\mathcal F}({\Lambda_t})$.
 
The following proposition represents the essence of dynamics
  (see also {\cite{Berthelot,Newton,Euler1,27}}) (in what follows, for the sake of brevity, we will not consider thermodynamics which, along with the motion equation, defines more fully the concept of mechanical system  \cite {Truesdell,Pobedria}).
  
{\bf Fundamental principle of dynamics}. 
{\em For a mechanical tube ${\bf \Lambda}\subset \tilde{\bf \Lambda}$  there exist  a Cartesian frame $\mathcal E_0$ and a parameterization ${\mathbf T}$ of the $\tau$--image such that the vector fields $ \vec {r} _ {x}$ and $\vec {v} _ {x}$ are solutions of 
 the following 
\begin{equation}
		\frac{d}{dt}{\mathcal P}^0({\Lambda_t})=
		{\mathcal F}^0({\Lambda_t})
		,\quad {\Lambda}_t\subset \tilde{\Lambda}_t, \ t\in {\mathbf T}\label{ 411}\vspace{-7pt}\end{equation}
{\rm In this case}
\begin{enumerate}{\parskip -0.05cm \vspace{-7pt}
\item the frame ${\cal E}_0$ and the parameterization ${\mathbf T}$ are called {\em inertial} (the frame is also called that of {\em reference} );}{\vspace{-7pt}
\item the aggregate $\alpha = \{\sigma _3$, $ \sigma _t$, $\mu$, $\forall t \in {\bf T}$, 
${\Lambda_t}\subset \tilde{\Lambda}_t$, $ {\mathcal P}({\Lambda_t})$, $ {\mathcal F}({\Lambda_t})\}$ is called {\em mechanical system};\vspace{-7pt}
 \item the set $\Lambda_t$ is called {\em (actual) shape} undergone by the mechanical system at $t \in {\bf T}$; \vspace{-7pt}
 \item the differentiable map ${\bf T}\rightarrow \{{\Lambda}_t$, $t \in {\bf T}
\}$ is called {\em motion} of the mechanical system \cite{Arnold};\vspace{-7pt}
 \item relation (\ref{ 411}) is called {\em motion equation};\vspace{-7pt}
 \item the integral ${\mathcal K}({\Lambda_t})$ is called {\em scalar measure of motion} of the mechanical system;\vspace{-7pt}
\item the screw ${\mathcal P}({\Lambda_t})$ is called {\em vector measure of motion} of the mechanical system;\vspace{-7pt}
\item the screw ${\mathcal F}({\Lambda} _t)$ is called {\em screw measure of impressed action} of the mechanical system $\alpha^e = \{\sigma _3$, $ \sigma _t$, $\mu$, $\forall t \in {\bf T}$, ${\Lambda}_t^e\subset \tilde{\Lambda}_t$,  $ {\mathcal P}({\Lambda} _t^e)$, $ -{\mathcal F}({\Lambda} _t)\}$ on the mechanical system $\alpha $  (it defines the action of the environment ${\bf \Lambda}^e$ on  ${\bf \Lambda}$).} \end{enumerate} 
}

Relation  (\ref{ 411}) can be transformed in the vector form 
\begin{equation}
	\frac{d}{dt}{\mathcal P}({\Lambda_t})={\mathcal F}({\Lambda_t})\label{ 4}
\end{equation}

Note that the parameterization and the frame ${\cal E}_0$ are elements of {\em Galilean group} of transformations of ${\bf A}_4$ which preserve intervals of time and the distance between simultaneous events \cite{Arnold}.

The set $\tilde{\Lambda}^{c}_t \subset \tilde{\Lambda}_t$ is called {\em set of concentration} of the measure ${\mathcal M}$ if ${\mathcal M}(\Lambda_t)=0$ for any set $\Lambda_t \subset \tilde{\Lambda}_t\setminus \tilde{\Lambda}^{c}_t$. We shall assume that relation (\ref{ 4}) is true only for ${\Lambda_t}\subset \tilde{\Lambda}^{c}_t$, $ t \in {\bf T}$. 
 \medskip
 
{\bf 1.3.   Concept of body}

The concept of a body is the subject of various mathematical formalizations. 
For example, one may represent a body as a point--wise set, an element of Boolean algebra, a differentiable manifold, a topological or measure space \cite{Truesdell,Konoplev1999,Pobedria} where a map into the space of shapes is considered. But there is a small obstacle: we must also transfer masses and forces to body shares.

If we do it in some way then the mathematical abstraction -- body with mass and force -- loses the primitive nature. In order to work out a mathematical theory of  mechanics we have all the necessary: shares kinematical and dynamic structures  attributed by them. 
 That is why (if it is necessary) we shall use the following conventions in the case of a given mechanical system 
$\alpha = \{\sigma _3$, $ \sigma _t$, $\mu$, $\forall t \in {\bf T}$, 
${\Lambda_t}\subset \tilde{\Lambda}_t$, $ {\mathcal P}({\Lambda_t})$, $ {\mathcal F}({\Lambda_t})\}$:
\begin{enumerate}{\em \parskip -0.25cm \vspace{-5pt}
\item \thinspace \thinspace the body is that takes some shapes ${\Lambda_t}\subset \tilde{\Lambda}_t$ in 3--dimensional affine space at some instants of time (cf. Aristotel, Physics, III, 5, 204b);
	\item \thinspace \thinspace 
	the change of shape undergone by a body from one instant to another is called the motion of that body (due to the principle of determinacy);
	\item \thinspace \thinspace the positive number ${\mathcal M}(\Lambda_t)$ is the body mass;
	\item \thinspace \thinspace the screw ${\mathcal F}({\Lambda_t})$ is the force 	impressed at the body.}\vspace{-7pt}
\end{enumerate}
Figuratively speaking, it is just what we see on the movie screen, achieved by a range of consecutive film shots \cite{Pobedria}.

Note that according to Glossary, Earth Observatory, NASA: force is any external agent that causes a change in the motion of a free body, or that causes stress in a fixed body.

While the concept of a mechanical system has strict mathematical sense, the concept of a body given above has only descriptive character, being a tribute of very fruitful tradition.\medskip

{\bf 1.4.  Generalization of mechanical system concept}
 
The non--trivial nature of the mechanical system concept can be seen from the fact that we may postulate the following equation of motion (instead of (\ref{ 4}))\vspace{-3pt}
\begin{equation}
\frac{d}{dt}{\mathcal P}({\Lambda_t})={\mathcal F}({\Lambda} _t)+{\mathcal F}_{\hskip -.05cm i}({\Lambda} _t)+{\mathcal F}_{\hskip -.05cm c}({\Lambda} _t)
	\label{ 7}
\end{equation}where 
\begin{equation}
	{\mathcal P}({\Lambda_t})\stackrel{{def}}{=}\int \hspace{-.05cm} \chi _{\hspace {-.05cm}_{ \Lambda_t}} l^{\hspace{.02cm}p_x, q_x} \mu (dx)\vspace{-3pt}
	\label{ 72}
\end{equation}
is the inhomogeneous {\em screw measure of motion} of the mechanical system 
(it is not necessary to think that    $\vec p=\rho_x \vec v_x $ in (\ref{ 72}));
${\mathcal F}$ is the inhomogeneous screw measure; the inhomogeneous screw measure ${\mathcal F}_i$ is so called {\em increment velocity} of the measure ${\mathcal P}$; the inhomogeneous screw measure ${\mathcal F}_c$ is so called {\em constraint action}. 

Assume that ${\mathcal F}_c={\mathcal F}_{int}+{\mathcal F}_{ext}
$ where ${\mathcal F}_{int}$ is formed by internal constraints of the set ${\Lambda_t}$ while ${\mathcal F}_{ext}$ is formed by external constraints.

Note that the base point in mechanics is that the motion of bodies is caused by interaction with their environment in the universe $\tilde{\bf \Lambda}$ which is described in the terms of force, moment of force and `innate' moment \cite{Zhilin}. However one must pay attention to constraints on bodies and their parts as well as interchange of masses, linear and angular momentums.
\medskip

{\bf 1.5. Derivatives of some measures}

Let us present the set $ {\Lambda_t} $ as the union of the set $ \Lambda_t^{pp}$ of the pure points entering into it, with the set $\Lambda_t^{ac}$ of its continuous points. We will assume that the last set has the surface $\partial \Lambda_t^{ac}$ which is Lyapunov's simple closed one \cite{Zhilin}.

Due to Gauss--Ostrogradsky (divergence) theorem, we have (see also \cite{Pobedria})
\[\frac{d}{dt}{\mathcal M}({\Lambda_t})=
\int \hspace{-.05cm} \chi _{\hspace {-.05cm}_{\Lambda_t^{ac}}}(\frac{d}{dt}\rho _x+\rho _x{\rm div}\hskip .05cm \vec v_x)\mu_{ac}(dx)+\sum_k (\frac{d}{dt}\rho _{x_k})\mu_{pp}(x_k)
\vspace{7pt}\] 
We shall assume that the function $\rho _x$ is defined by 
the {\em continuity equation} for continuous points 
$
\frac{\partial}{\partial t}\rho _x+{\rm div}\hskip .02cm \vec p_x=\nu _x
$ and for pure points -- 
$
\frac{d}{dt}\rho _{x_k}=\nu _{x_k}
$ where $\nu _x$ depicts the generation (negative in the case of removal) per unit volume and unit time of the measure ${\mathcal M}$. Terms that generate ($\nu _x> 0$ ) or remove ($\nu _x< 0$) are referred to as `sources' and `sinks' respectively. 

In what follows, we shall assume that all vector and tensor sliders are homogeneous. 

According to  \cite{Pobedria}
\begin{equation}
	\frac{d}{dt}{\mathcal P}(\Lambda_t)=
	\int \hspace{-.05cm} \chi _{\hspace {-.05cm}_{\Lambda_t^{ac}}}(\frac{d}{dt}
	l^{\hskip .02cm p_x}+
	l^{\hskip .02cm p_x}
	{\rm div}\hskip .05cm \vec v_x)\mu_{ac}(dx)+\sum_k (\frac{d}{dt}l^{\hskip .02cm p_{x_k}})\mu_{pp}(x_k)
	 \label{ 5}\vspace{3pt}
\end{equation}
As for continuous points (see also \cite{Pobedria})
\[\frac{d}{dt}
l^{\hskip .02cm p_x}+
l^{\hskip .02cm p_x}
{\rm div}\hskip .05cm \vec v_x=
\rho _x
\frac{d}{dt}\hskip .05cm l^{\hskip .02cm {v}_x}+\nu _xl^{\hskip .02cm {v}_x}\]
from (\ref{ 7}) and (\ref{ 5}) follows 
\begin{equation}
\int \hspace{-.05cm} \chi _{\hspace {-.05cm}_{{\Lambda_t}}}(
\rho_x
\frac{d}{dt}\hskip .05cm 
l^{\hskip .02cm {v}_x}+\nu _x l ^{\hspace{.02cm}{v}_x})\mu(dx)
={\mathcal F}({\Lambda} _t)+{\mathcal F}_i({\Lambda} _t)+{\mathcal F}_c({\Lambda} _t)
\label{ 6} \end{equation} 
\newpage

{\large \bf 2.  Specifying mechanical systems}

Show how the given above axiomatics relates to the conventional mechanics. 
 In the first place exemplify the notion of skew  screw bi--measure. In the conventional mechanics it is considered that there is the gravitational interaction between bodies. It can be formalized in the following way. Let a skew  screw bi--measure 
${\it \Psi}({\Lambda} _t, \Lambda_t^{e})$ be  such that 
\[{\it \Psi}({\Lambda} _t, \Lambda_t^{e}) =\int \hspace{-.05cm} \chi _{{\Lambda} _t} l^{g_x} \rho_x \mu(dx),\quad \vec g_x=\gamma 
\int \hspace{-.05cm} \chi _{\hspace {-.05cm}_{\Lambda_t^{e}}} 
\overrightarrow{ (x-y)}\hspace{.1cm}
\frac{\rho_y \mu(dy)}{\Vert
\overrightarrow{ (x-y)}\Vert ^3}\vspace{7pt}
\]
where $\gamma $ is a positive (gravitational) constant, the $\mu $-integrable homogeneous  slider $\rho_x l^{\hspace{.02cm} g_x}$ is defined at $x\in {\Lambda}_t$. 

Then the screw  ${\mathcal G}({\Lambda} _t)\stackrel{\rm { def}}{=} {\it \Psi}({\Lambda} _t, \Lambda_t^{e})$ can be called 
{\em measure of gravitating action} of $\alpha^e$ upon $\alpha$ {\rm \cite{Konoplev1999}}.
One may take this screw as the screw measure ${\mathcal F}({\Lambda} _t)$  of impressed action.

Assume that the increment velocity of ${\mathcal P}$ is given by the following Lebesgue--Stieltjes integral
\begin{equation}
	{\mathcal F}_i({\Lambda_t})=\int \hspace{-.05cm} \chi _{\hspace {-.05cm}_{ \Lambda_t}} l^{\hspace{.02cm}\xi _x}\hspace{.005cm} \mu (dx), \quad \Lambda_t\subset \tilde{\Lambda}_t \label{fi}\vspace{-3pt}
\end{equation}where $l^{\hspace{.02cm}\xi _x}$ is its density.

Let no external constraint be.\medskip

{\bf 2.1. A mass--point} 

Consider a world--line ${\bf \Lambda}\subset \tilde{\bf \Lambda}$ whose image in ${\bf A}_3\times {\bf T}$ generates the curve $\{{x }(t)\in \tilde{\Lambda}_t, t \in {\bf T}\}$. Assume that the points $x(t)$ are pure, {\em i.e.}, $x(t)=x_k(t)$. Then the mechanical system $\alpha = \{\sigma _3$, $ \sigma _t$, $\mu$, $\forall t \in {\bf T}$, 
$x(t)\in \tilde{\Lambda}_t$, $\rho _x$, $\nu _x$, $ \vec f_x$, $\vec \xi _x\}$ is called {\em mass--point}. 

From relation (\ref{ 6}) follows that 
\begin{equation}\rho _x
\frac{d}{dt}\hskip .05cm \vec v_x+\nu _x\vec v_x=\vec f_x+\vec \xi _x 
\label{ 8}
\end{equation}
If $\nu _x\equiv0$ and $\vec \xi _x\equiv0$, then equation (\ref{ 8})
is known as {\em second Newton's law} where $\vec { f}_x$ is the impressed force acting at the point $x=x_k\in \tilde{\Lambda}_t$ with the mass ${\mathcal M}_k=\rho _x \mu_k$. 

If $\nu _x\neq 0$ and $\vec \xi _x=\nu _x\vec u_x$ where $\vec u_x$ is the velocity of mass gain or loss, then equation (\ref{ 8}) is known as that of Meshchersky \cite{27}. 
\begin{Remark}A classical example of mass--points with constraints is the mechanical system  khown as pendulum.
\end{Remark}\newpage

{\bf 2.2. A rigid body}

The mechanical system 
$\alpha_p = \{\sigma _3$, $ \sigma _t$, $\mu$, $\forall t \in {\bf T}$, 
${\Lambda_t}\subset \tilde{\Lambda}_t$, $\forall x\in {\Lambda_t}$, $\rho _x$, $\nu _x$, $ \vec f _x$, 
$ \vec \xi _x\}$ is called {\em rigid body} if
\begin{enumerate}{\em \parskip -0.1cm \vspace{-7pt}
\item the sets $\Lambda_t$ are bounded and closed;
\item the constraints applied on its points keep distances between them not changing with time;
\item the constraints are {\em ideal} \cite{Vilke}.}\vspace{-7pt}
\end{enumerate}
A rigid body may contain continuous and pure points \cite{27}.
\medskip

{\bf 2.2.1.  Newton--Euler equation}. 
At any time instant $t^{\ast}$ consider the set ${\Lambda}_{t^{\ast}}$. Let  a Cartesian frame ${\cal E}_p$ be attached to the set under consideration. It is plain that the frame takes the same position in all sets ${\Lambda}_t$. In the frame these sets are immobile, coincide one with another and form the set noted as ${\Lambda}_p$ in the frame ${\cal E}_p$. We shall say that the frame ${\cal E}_p$ is attached to the rigid body $\alpha_p$.
 
Let us bound the vectors $ \vec v _{0,p}$ and $ \vec \omega  _{0,p}$ at the point $O_p$ and 
the vectors $ \vec v _x$ and $ \vec \omega  _{0,p}$ at points $x\in {\bf E}_3$. Then due to (\ref{cross}) we have the field $V_{0,p}=l^{\omega  _{0,p},v _{0,p}}=\{\vec \omega_{\hspace{.01cm}{0},{p}},\vec v_{\hspace {.01cm}0,p} + \vec r_{\hspace {.01cm}x,p}{\times}\vec \omega _{\hspace {.01cm}0,p}, \forall x\in {\bf E}_3\}$  known as {\em kinematic slider}. The coordinate representation of its reduction (twist) $V^{tw,p}_{0,p}={\rm col}\{v_{0,p}^p,\omega _{0,p}^p\}\in {\bf R}_6$ is called {\em vector of quasi-velocities}. 
\begin{Lemma}
There is the following relation {\rm \cite{Konoplev1999}}
\[
{l}_{p}^{v_x,wr,p}={\it \Theta }_p^xV^{tw,p}_{0,p},\quad {\it \Theta }_p^x= 
\begin{bmatrix}
I & -{r}_{p,x}^{p\times} \\ 
 r_{p,x}^{p\times} & -({r}_{p,x}^{p\times} )^2
\end{bmatrix}
\]
\end{Lemma}
\proof 
The statement is true as \vspace{5pt}
\[
{l}_p^{v_x,wr,p} =
\begin{bmatrix}
I \\ 
 {r}_{p,x}^{p\times}
\end{bmatrix} v_x^p=
\begin{bmatrix}
I \\ 
 {r}_{p,x}^{p\times} 
\end{bmatrix} \left(v_{0,p}^p - {r}_{p,x}^{p\times} \omega_{0,p}^p\right) = 
\begin{bmatrix}
I & -{r}_{p,x}^{p\times} \\ 
 {r}_{p,x}^{p\times} & -({r}_{p,x}^{p\times} )^2
\end{bmatrix} 
\begin{pmatrix}
v_{0,p}^p \\ 
\omega_{0,p}^p
\end{pmatrix}
\vspace{5pt}\]
According to the rigid body definition the internal constraints are considered as ideal and thus \cite{Vilke} 
	\[{\mathcal F}_{\hskip -.05cm c}({\Lambda} _t)=0
\]
From relations  (\ref{ 6})--(\ref{fi}) follows that \cite{Pobedria}\vspace{-3pt}
\begin{equation}
	\rho_x
	\frac{d}{dt}\hskip .05cm 
	l_0^{\hskip .02cm  {v}_x,0}+\nu _x l_0 ^{\hspace{.02cm}{v}_x,0}=l_0^{\hskip .02cm  {f}_x,0}+l_0^{\hskip .02cm  \xi _x,0}\label{body}\vspace{-3pt}
\end{equation}
Hence  we have
\[\int \hspace{-.05cm} \chi _{\hspace {-.05cm}_{{\Lambda}_p}}[
\rho _x { L}_{0,p}^{-1}
	\frac{d}{dt}\hskip .05cm ({ L}_{0,p}
	{\it \Theta }_p^x V_{0,p}^{tw,p})+\nu _x 
	{\it \Theta }_p^x V_{0,p}^{tw,p} ]\mu(dx)=\int \hspace{-.05cm} \chi _{\hspace {-.05cm}_{{\Lambda}_p}}[
	l_p^{\hskip .02cm  {f}_x,p}+l_p^{\hskip .02cm  \xi _x,p}
 ]\mu(dx)
\]  
As the twist $V_{0,p}^{tw,p}$ does not depend on points $x$, the following statement is true.
\begin{Theorem} 
The motion of $\alpha_p$ (w.r.t. ${\cal E}_0$ in the frame ${\cal E}_p$) is
described by the ({\em Newton--Euler}) equation {\rm \cite{Konoplev1999}} 
\begin{equation}{ \it \Theta }_p V_{0,p}^{tw,p \centerdot } +(Q_p+\varPhi^{wr}_{0,p}{ \it \Theta }_p )V^{tw,p}_{0,p} ={\mathcal F}_{\hskip -.05cm p}^{wr,p}({\Lambda}_p)
+{\mathcal F}_{\hskip -.05cm i\hskip .005cm p}^{wr,p}({\Lambda}_p)
\label{ 6g}
\end{equation}
where $Q_p=\hspace{-0.1cm} \int \hspace{-.05cm} \chi _{\hspace {-.05cm}_{{\Lambda}_p}} {\it \Theta }_p^x \nu_x \mu(dx)$, ${ \it \Theta }_p=\hspace{-0.1cm} \int \hspace{-.05cm} \chi _{\hspace {-.05cm}_{{\Lambda}_p}} {\it \Theta }_p^x \rho _x \mu(dx)$.
\end{Theorem}
\medskip

{\bf 2.2.2. Systems of consecutively connected rigid bodies { \cite{Konoplev2010}}}. Let us consider a system of $k+1$ consecutively connected rigid bodies $\alpha_p$, $p=\overline{0,k}$ (the rigid body $\alpha_0$ is immobile). Its motion is depicted by the following Newton--Euler equation
\begin{equation}
	A V_a^{{}^ \centerdot}+B V_a= F_a
	\label{ne}
\end{equation}
where $A$ and $B$ are known matrices, $V_a={\rm col}\{V^{tw,p}_{0,p}\}$, $F_a={\rm col}\{{\mathcal F}_{p}^{wr,p}+{\mathcal F}_{\hskip -.05cm i\hskip .005cm p}^{wr,p}\}$, $p=\overline{1,k}$. 

 Newton--Euler equation (\ref{ne}) is considered w.r.t. `absolute' quasi--velocities $V^{tw,p}_{0,p}$ of the rigid bodies (calculated in ${\mathcal E}_p$ w.r.t. the main frame ${\mathcal E}_0$). But in practice there are only the `relative' quasi--velocities $V_{p-1,p}^{tw,p}$ of the frame ${\mathcal E}%
_p$ w.r.t. ${\mathcal E}_{p-1}$. 
 Thus we must connect the `absolute' quasi--velocities with `relative' ones.
\begin{Lemma} 
For a system of consecutively connected rigid bodies there is the following composition rule {\em \cite{Konoplev1999}} 
\begin{equation}V^{tw,p}_{0,p}=\sum_{s=1}^{s=p}L_{p,s}^{tw}V_{s-1,s}^{tw,s}, \quad V_{s-1,s}^{tw,s}= 
\begin{pmatrix}
v_{s-1,s}^{s} \\ 
\omega _{s-1,s}^{s}
\end{pmatrix}
\label{ 16}\vspace{7pt}
\end{equation}
where ${ L}_{p,s}^{tw}$ is given as in (\ref{L^{tw}}) for ${ L}^{wr}_{p,s}={C}_{p,s}^\otimes { D}^s_{p,s}=({ D}^s_{s,p}{C}_{s,p}^\otimes)^{-1}={ L}_{s,p}^{wr,T}$.
\end{Lemma}
\proof 
As the rigid bodies are connected consecutively there is the relation 
${ C}_{0,p}={ C}_{0,s}{ C}_{s,p}
$. With differentiating it we have 
 $\omega _{0,p}^p=\omega _{0,s}^p+\omega
_{s,p}^p=\omega _{0,s}^p+C_{p,s}\omega _{s,p}^s
$. 

Let us define the vectors $\vec {d}_{s,p}=\overrightarrow{O_sO_p}$ and $\vec {d}_{0,s}=\overrightarrow{O_0O_{s}}$, then $\vec {d}_{0,p}=\overrightarrow{O_0O_p}=\vec {d}_{0,s}+\vec {d}_{s,p}$, 
$v_{0,p}^p=v_{0,s}^p+d_{s,p}^{p\centerdot }$, $d_{s,p}^p=C_{p,s}d_{s,p}^{s}$, $d_{s,p}^{p\centerdot
}=v_{s,p}^p+C_{p,s}^{{}^\centerdot }d_{s,p}^{s}=v_{s,p}^p-\omega _{s,p}^{p\times
}d_{s,p}^p=v_{s,p}^p+d_{s,p}^{p\times }\omega _{s,p}^p$. Hence $v_{0,p}^p=v_{0,s}^p+C_{p,s}d_{s,p}^{s\times
}\omega _{s,p}^{s}+v_{s,p}^p$, and 
\[
V^{tw,p}_{0,p}=V_{0,s}^{tw,p}+{ L}_{p,s}^{tw}V_{s,p}^{tw,s}\vspace{-5pt}\] 
Hence we have (\ref{ 16}).

From (\ref{ 16}) follows the {\em equation of kinematics}\index{Equation!of kinematics} 
\begin{equation}
{ V}_a={ L} { V}_{r}
\label{ 17}
\end{equation}
where ${V}_a\hspace{-.1cm}=\hspace{-.1cm}{\rm  col}%
\{{\it V}_{0,1}^{tw,1},\ldots ,\hspace{-.1cm} {\it V}_{0,\it p}^{tw,p},\ldots ,\hspace{-.1cm}{\it V}_{ 0,{\it k}}^{tw,k}\}$, ${ V}_r\hspace{-.1cm}=\hspace{-.1cm}{\rm  col}%
\{{\it V}_{0,1}^{tw,1},\ldots ,\hspace{-.1cm}{\it V}_{{\it p}-1,\it p}^{tw,p},\ldots ,\hspace{-.1cm}{\it V}_{ {\it k}-1,{\it k}}^{tw,k}\}$, ${ L}$ is the triangular matrix with blocks ${ L}_{p,s}^{tw}$ being functions of `relative' frame rotations and translations (and their velocities).

Thus we have 
\begin{equation}
A L V_r^{\hspace{.02cm }{}^\centerdot}+(A L^{\hspace{.02cm }{}^\centerdot} +B)V_r= F_a \label{NewtonE}
\end{equation}
where $L^{\hspace{.02cm }{}^\centerdot}$ is analytically calculated due to relation (\ref{ 41}).

It is easy to see that the matrices of relation (\ref{NewtonE}) depend on rotation matrices (and linear and angular quasi--velocities, too) that is why equation (\ref{NewtonE}) must be considered along with the {\em Euler kinematical relation}
\begin{equation}
 C_{p-1,p}^{\hspace{.02cm }{}^\centerdot}=C_{p,p-1}\omega _{p-1,p}^{p\times}
\label{Ek}
\end{equation}
\vspace{7pt}

\begin{figure}[ht]
\begin{center}
\hspace{-.06cm}\includegraphics[width=9truecm, viewport=0 0 1416 759]{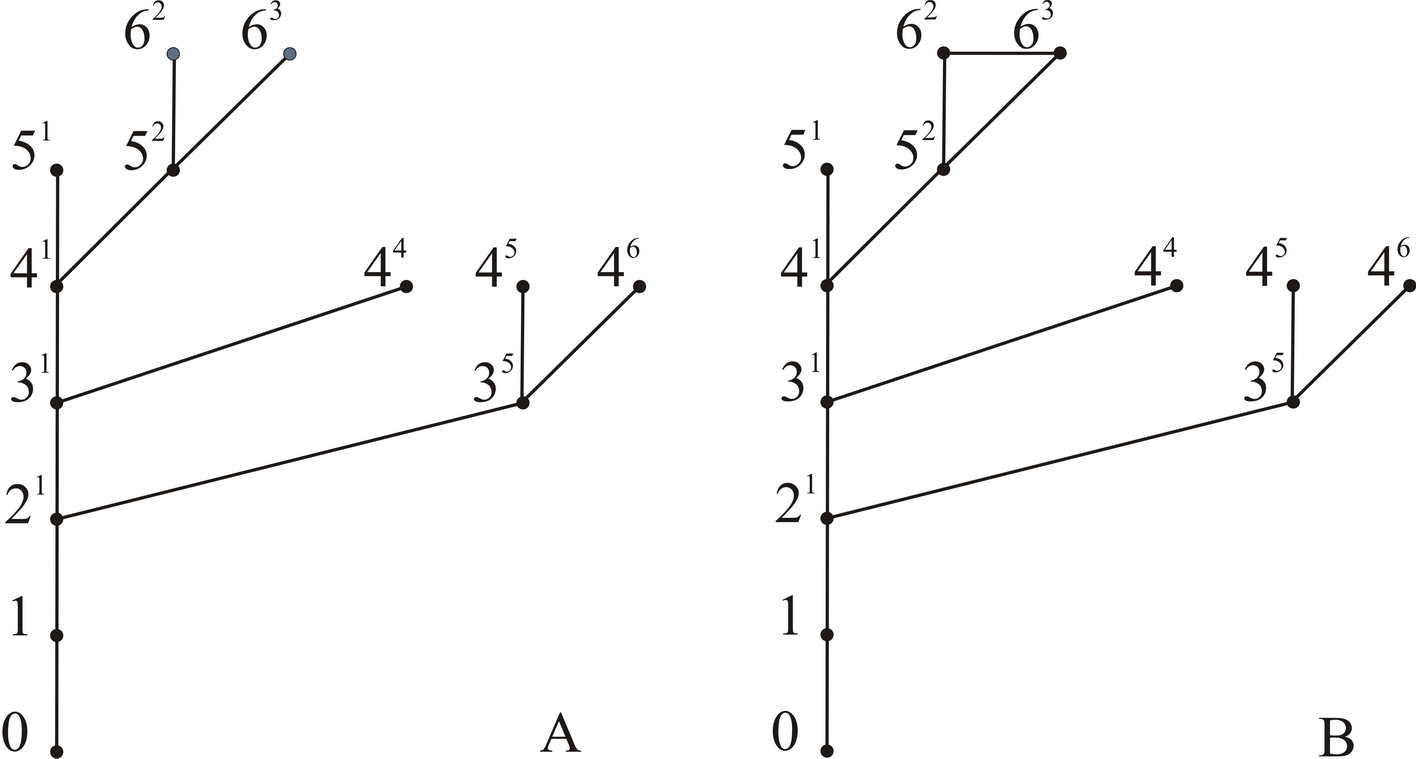}
\end{center}\vspace{-7pt}
\begin{center}%
{Fig. 1. Multibody system graphs}\end{center}
 \vspace{-.5cm} 
 \end{figure} 
{\bf 2.2.3.  Multibody systems with tree--like structure}.
Consider a multibody system with tree--like structure given by 
 the graph in Fig. 1A.
 Let vertices $j^i$ represent the system bodies or the origins of the attached Cartesian frames ${{\mathcal E}}_j^i$ where the index $i$ numbers the tree--tops, the index $j$ numbers the rigid bodies from the base to the corresponding tree--tops. Introduce $V_{j,p}^{m,i}$ as quasi--velocities characterizing rotation and translation of the frames ${{\mathcal E}}_j^i$ w.r.t. ${{\mathcal E}}_p^m$. Then we have the sets 
 $\{V_{0,\hspace{-.01cm}1}^{0,\hspace{-.01cm}1},\hspace{-.01cm} V_{1,\hspace{-.01cm}2}^{1,\hspace{-.01cm}1},\hspace{-.01cm} V_{2,\hspace{-.01cm}3}^{1,\hspace{-.01cm}1},\hspace{-.01cm} V_{3,\hspace{-.01cm}4}^{1,\hspace{-.01cm}1}, V_{4,\hspace{-.01cm}5}^{1,\hspace{-.01cm}1}\},\hspace{-.01cm}
\{V_{0,\hspace{-.01cm}1}^{0,\hspace{-.01cm}1},\hspace{-.01cm} V_{1,\hspace{-.01cm}2}^{1,\hspace{-.01cm}1},\hspace{-.01cm} V_{2,\hspace{-.01cm}3}^{1,\hspace{-.01cm}1}, V_{3,\hspace{-.01cm}4}^{1,\hspace{-.01cm}1},\hspace{-.01cm}
V_{4,\hspace{-.01cm}5}^{1,\hspace{-.01cm}2}, V_{5,\hspace{-.01cm}6}^{2,\hspace{-.01cm}2}\},\hspace{-.01cm}
\{V_{0,\hspace{-.01cm}1}^{0,\hspace{-.01cm}1},$
 
\noindent 
$
V_{1,\hspace{-.01cm}2}^{1,\hspace{-.01cm}1},
V_{2,\hspace{-.01cm}3}^{1,\hspace{-.01cm}1}, V_{3,\hspace{-.01cm}4}^{1,\hspace{-.01cm}1},
V_{4,\hspace{-.01cm}5}^{1,\hspace{-.01cm}2}, V_{5,\hspace{-.01cm}6}^{2,\hspace{-.01cm}3}\},\
\{V_{0,\hspace{-.01cm}1}^{0,\hspace{-.01cm}1} V_{1,\hspace{-.01cm}2}^{1,\hspace{-.01cm}1}, V_{2,\hspace{-.01cm}3}^{1,\hspace{-.01cm}1}, V_{3,\hspace{-.01cm}4}^{1,\hspace{-.01cm}4}\},\
\{V_{0,\hspace{-.01cm}1}^{0,\hspace{-.01cm}1} V_{1,\hspace{-.01cm}2}^{1,\hspace{-.01cm}1} V_{2,\hspace{-.01cm}3}^{1,\hspace{-.01cm}5} V_{3,\hspace{-.01cm}4}^{5,\hspace{-.01cm}5}\},\
\{V_{0,\hspace{-.01cm}1}^{0,\hspace{-.01cm}1}, V_{1,\hspace{-.01cm}2}^{1,\hspace{-.01cm}1}, V_{2,\hspace{-.01cm}3}^{1,\hspace{-.01cm}5},
V_{3,\hspace{-.01cm}4}^{5,\hspace{-.01cm}6}\}$ 
and $\{V_{0,1}^{0,1}, 
V_{0,2}^{1,1}\hspace{-.01cm},\hspace{-.01cm} V_{0,3}^{1,1}\hspace{-.01cm},\hspace{-.01cm} V_{0,4}^{1,1}, V_{0,5}^{1,1}\},
\{V_{0,1}^{0,1}\hspace{-.01cm},\hspace{-.01cm} V_{0,2}^{1,1}\hspace{-.01cm},\hspace{-.01cm}V_{0,3}^{1,1}\hspace{-.01cm},\hspace{-.01cm} V_{0,4}^{1,1}\hspace{-.01cm},\hspace{-.01cm}V_{0,\hspace{-.01cm}5}^{1,\hspace{-.01cm}2}\hspace{-.01cm},\hspace{-.01cm}
V_{0,\hspace{-.01cm}6}^{2,\hspace{-.01cm}2}\},
\{V_{0,\hspace{-.01cm}1}^{0,\hspace{-.01cm}1}\hspace{-.01cm},\hspace{-.01cm}V_{0,\hspace{-.01cm}2}^{1,\hspace{-.01cm}1}\hspace{-.01cm},\hspace{-.01cm} V_{0,\hspace{-.01cm}3}^{1,\hspace{-.01cm}1}\hspace{-.01cm},\hspace{-.01cm} V_{0,\hspace{-.01cm}4}^{1,\hspace{-.01cm}1}\hspace{-.01cm},\hspace{-.01cm} V_{0,\hspace{-.01cm}5}^{1,\hspace{-.01cm}2}\hspace{-.02cm},$

 \noindent 
$\hspace{-.01cm}V_{0,\hspace{-.01cm}6}^{2,\hspace{-.01cm}3}\},\{V_{0,\hspace{-.01cm}1}^{0,\hspace{-.01cm}1}\hspace{-.01cm}, 
V_{0,\hspace{-.01cm}2}^{1,\hspace{-.01cm}1}\hspace{-.01cm},\hspace{-.01cm} V_{0,\hspace{-.01cm}3}^{1,\hspace{-.01cm}1}\hspace{-.02cm},\hspace{-.01cm} V_{0,\hspace{-.01cm}4}^{1,\hspace{-.01cm}4}\},
\{V_{0,\hspace{-.01cm}1}^{0,\hspace{-.01cm}1}\hspace{-.01cm},\hspace{-.01cm} V_{0,\hspace{-.01cm}2}^{1,\hspace{-.01cm}1},V_{0,\hspace{-.01cm}3}^{1,\hspace{-.01cm}5}\hspace{-.01cm},\hspace{-.01cm}
V_{0,4}^{5,5}\}\hspace{-.01cm},\hspace{-.01cm}
\{V_{0,1}^{0,1}\hspace{-.02cm},\hspace{-.01cm}V_{0,2}^{1,1}\hspace{-.01cm},\hspace{-.01cm} V_{0,3}^{1,5}\hspace{-.01cm},\hspace{-.01cm} V_{0,4}^{5,6}\}$
with the same subscripts as in the case of consecutively connected rigid bodies for the relative and absolute quasi--velocities. 
 This case is considered above that is why we arrive at relation 
(\ref{ 17}) with the known matrix ${L}$ and $
{ V}_a = {\rm  col}\{
V_{0,1}^{0,1}, V_{0,2}^{1,1}, V_{0,3}^{1,1}, V_{0,4}^{1,1}, V_{0,5}^{1,1}
, V_{0,5}^{1,2}, V_{0,6}^{2,2}
,V_{0,6}^{2,3},V_{0,4}^{1,4},
V_{0,3}^{1,5}, V_{0,4}^{5,5}
,V_{0,4}^{5,6}\}$, $
{ V}_r  =
{\rm  col}\{
V_{0,1}^{0,1}, V_{1,2}^{1,1}, V_{2,3}^{1,1}, V_{3,4}^{1,1}, V_{4,5}^{1,1}, 
V_{4,5}^{1,2}, V_{5,6}^{2,2},
V_{5,6}^{2,3},V_{3,4}^{1,4},
V_{2,3}^{1,5}, V_{3,4}^{5,5},
V_{3,4}^{5,6}\}$.
 \begin{Remark}The results obtained can be immediately applied to systems with loops, {\em e.g.}, if in the system under consideration (see Fig. 1B) the vertex $6^2$ is connected with $6^3$ by the edge ($6^2,6^3$). In this case relation (\ref{ 6g}) is the same, but in the case where constraints are considered there are the following additional constraints $\overrightarrow{(5^2,6^2)}+\overrightarrow{(6^2,6^3)}+\overrightarrow{(6^2,5^2)}=0$ and $C^{2,2}_{5,6}C^{2,3}_{6,6}C^{2,2}_{6,5}=I$.
\end{Remark}

{\bf 2.2.4. Parameterization of rotation matrices}. 
The order of system (\ref{NewtonE})--(\ref{Ek}) may be reduced. To this end one uses different parameterizations of rotation matrices. 

{\bf 2.2.4.1.  Euler angles}. Let $C_{p-1,p}=C_1C_2C_3$ where\vspace{3pt} \begin{equation}
C_1 =
\begin{bmatrix}\hspace{.1cm}
1\ & 0 & 0 \\ \hspace{.1cm}
0\ & \cos \varphi & -\sin
\varphi \\ \hspace{.1cm} 
0\ & \sin \varphi & %
\cos \varphi 
\end{bmatrix} ,\, C_2 =
\begin{bmatrix}
\cos \vartheta &\ 0 & \sin \vartheta \\ 
0 &\ 1 & 0
\\ 
-\sin \vartheta &\ 0 & %
\cos \varphi 
\end{bmatrix} ,\, C_3 =
\begin{bmatrix}\hspace{.1cm}
\cos \psi & -\sin \psi \hskip %
-.04in &\ 0 \hspace{.1cm} \\ \hspace{.1cm} 
 \sin \psi & \cos \psi 
&\ 0 \hspace{.1cm} \\ \hspace{.1cm} 
0 & 0 &\ 1\hspace{.1cm}
\end{bmatrix}
\label{ 26}\vspace{3pt}
\end{equation}
are so called the simplest rotation\index{Rotation!simplest} matrices; $\varphi $, $\vartheta $, and $\psi $ are 
Euler angles \cite{Korn}.

Introduce the triple $ \lambda_{p-1,p}={\rm col}\{\varphi , \vartheta , \psi\}$ as a parameter. Then there is the matrix $D_{p-1,p}$ such that \cite{Konoplev1999}
\begin{equation}\omega _{p-1,p}^p=D_{p-1,p}\lambda_{p-1,p}^{{}^\centerdot}
\label{ 7g}\end{equation}

Hence equation (\ref{NewtonE}) must be considered along with the following relation
\begin{equation}\lambda_{{p-1,p}}^{{}^\centerdot}=D_{p-1,p}^{-1}\omega _{p-1,p}^p
\label{ 7ghh}\end{equation}
and $C_{p-1,p}=C_{p-1,p}(\lambda_{p-1,p})$ if the matrix $D_{p-1,p}^{-1}$ exists.\medskip

{\bf 2.2.4.2. Fedorov vector--parameter}. To parameterize rotation matrices we may introduce Fedorov vector--parameter \cite{Fedorov}.
\begin{Definition}{\rm \cite{Fedorov}} 
The number triple $f\in {\bf E}_3$ is called {\em Fedorov vector--parameter} of a rotation matrix $C$, if it corresponds to the following matrix 
\[
 f^{\times}= 
(C-I)(C+I)^{-1}
\]
\end{Definition}

The inverse map of Cayley restores the rotation matrix 
\[
 C{=} (I+f^{\times})(I-f^{\times})^{-1}
\]
It is easy to be verified (for example, by means of Maple$^ {\copyright} $) that 
 the following relations are true
\[
	f^{\times } =\frac{C-C^T}{1+{\rm tr}\hspace{0.05cm}\mathit{C}},
\quad
	C =\frac{(1-\Vert f \Vert ^2)I+2f f^T+2f^{\times }}{1+\Vert f \Vert ^2}
\]

Let the rotation matrices $C_{p,k}$ have Fedorov vector--parameter $f_{p,k}$. It is known that it is an eigenvector of $C_{p, k}$, {\em i.e.}, $C_{p,k}f_{p,k}=f_{p,k} \in {\bf E}_3
$. 

As the space ${\bf E}_3$ has 3 bases ${\bf e}_0$, ${\bf e}_p$ and ${\bf e}_k$ we may define the following vectors 
\[\vec g_{p,k}= \sum_i f_i \vec e_{0,i}, \quad 
\vec r_{p,k}= \sum_i f_i \vec e_{p,i}=\sum_i f_i \vec e_{k,i}
\]
where 
${\rm col}\{f_1, f_2, f_3\}=f_{p,k}$.
\begin{Definition}
The vector $\vec g_{p,k}$ is called vector of {\em Gibbs}, while 
$\vec r_{p,k}$ is called vector of {\em Rodrigues}.
\end{Definition}
 \begin{Remark}In \cite{Fedorov} it is explicitly pointed out that Fedorov vector--parameter $f_{p,k}$ is {\em Gibbs} vector $\vec g_{p,k}$, and $\vec f_{p,k}=\vec g_{p,k}$ as there is no other basis except the canonical one in \cite{Fedorov}. In general the vector $\vec g_{p,k}$ does not coincide with $\vec r_{p,k}$ (see Fig. 2), as the bases ${\bf e}_0$, ${\bf e}_p$ and ${\bf e}_k$ are different. Moreover the name of {\em vector} is used here conditionally as there is not the parallelogram rule for the vectors of the kind $\vec g_{p,k}$ and $\vec {r}_{p, k}$ (see also \cite{Zhilin,Fedorov, Lurie}
 ).  
\end{Remark}

\begin{figure}[ht] 
\begin{center}
\hspace{-.06cm}\includegraphics[width=8truecm, viewport=0 0 1417 683]{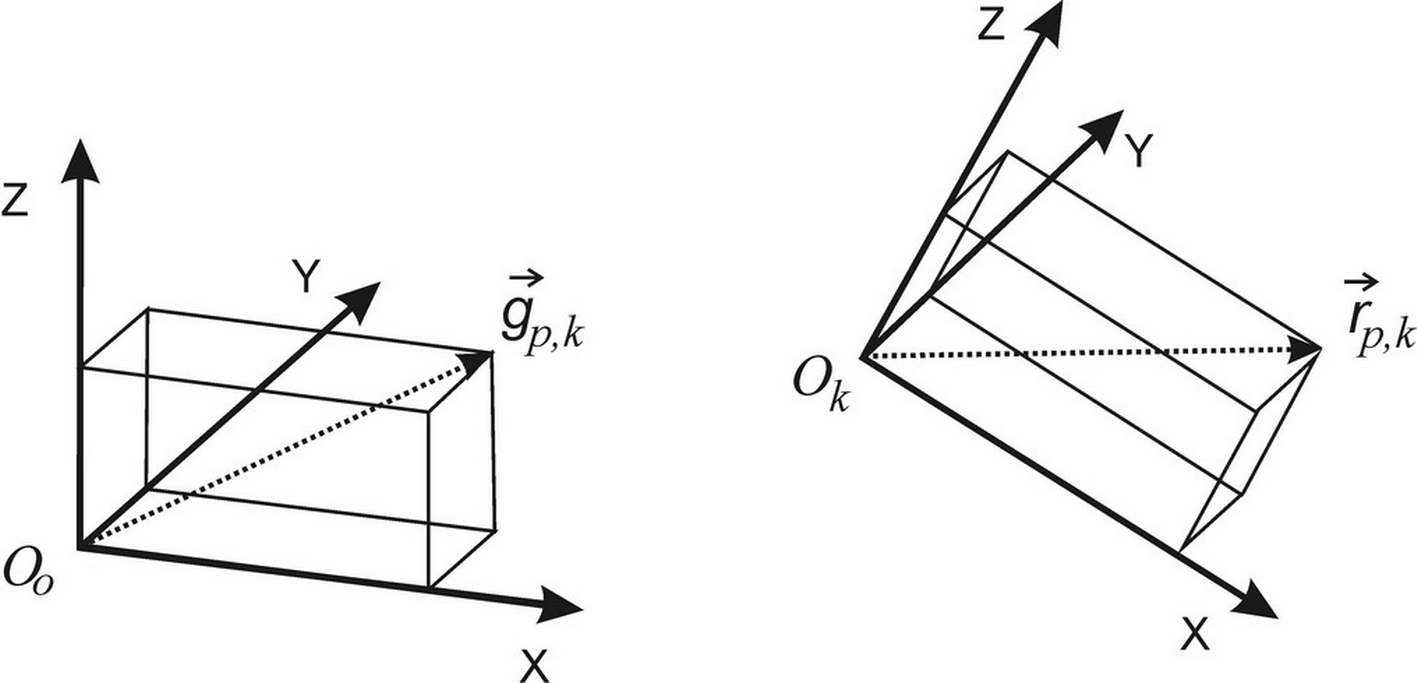}
\end{center}
\begin{center}%
{Fig. 2. Gibbs and Rodrigues vectors.}\end{center}
 \vspace{-.5cm} \end{figure} 

It is easy to see that the vector $\vec {r}_{p,k}$ is collinear with the instantaneous angular
velocity $\vec {\omega}_{p,k}$, and thus it defines the rotation axis ($\vec {\omega}_{p,k}$ is the half of the vector of finite rotation \cite{Lurie}).
  
There is the following relation \cite{Lurie} 
\[ 
\vec {\omega}_{p-1,p}=\frac 2{1+\Vert \vec {r}_{p-1,p}\Vert ^2}( \vec {r}^{\hspace{ .08cm}{}^\centerdot }_{p-1,p}+
\vec {r}_{p-1,p}\times \vec {r}^{\hspace{ .08cm}{}^\centerdot }_{p-1,p})
\]
where $\times $ means {\em vector product}.

 As the vectors $\vec {\omega}_{p-1,p}$ and $\vec {r}_{p-1,p}$ are collinear, we have relation (\ref{ 7g}) 
where $\lambda_{p-1,p}={f}_{p-1,p}$ and $D_{p-1,p}=\frac 2{1+\Vert {f}_{p-1,p}\Vert ^2}(I+
{f}_{p-1,p}^{\times})$.

Thus equation (\ref{NewtonE}) must be considered along with the following relation
\begin{equation}
f_{p-1,p}^{{}^\centerdot}=\frac 12\, (1+\Vert {f}_{p-1,p}\Vert^2)(I-f_{p-1,p}^{\times})^2(I+f^{\times}_{p-1,p})\omega _{p-1,p}^p
\label{ 7ghha}\end{equation}
and $C_{p-1,p}=C_{p-1,p}(f_{{p-1,p}})$.
\medskip 

{\bf 2.2.4.3. Euler--Rodrigues parameters}. To parameterize rotation matrices we may use quaternions. 
\begin{Definition}The set $\Lambda=\{{\lambda_0}\in {\bf R}, \vec \lambda\in {\bf V}_3\}$ is called {\em quaternion}.
\end{Definition} Quaternions generate the algebra with the quaternion product 
\[\Lambda \circ M=\{{\lambda_0}{\mu_0}-\langle \vec \lambda,\vec {\mu} \rangle , {\lambda_0}\vec \mu+{\mu_0}\vec \lambda+ \vec \lambda\times \vec \mu\}
\]
where $M=\{{\mu_0}, \vec \mu\}$.

Any vector $\vec \lambda$ can be imaged as a quaternion $\Lambda$ with the zero scalar part. That is why we may define the quaternion product of two vectors $\vec \lambda$ and $\vec {\mu}$ as follows 
\[\vec \lambda\circ \vec {\mu}=\{-\langle \vec \lambda,\vec {\mu} \rangle, \vec \lambda\times \vec \mu\}
\]
There exists the unit quaternion ${\Lambda}_{p-1,p}=\{{\lambda_0}, \vec \lambda_{p-1,p}\}$ (with $\Vert\Lambda_{p-1,p}\Vert = 1$) such that \cite{Yakovenko}
\begin{equation}
{\Lambda}_{p-1,p}\circ\vec \omega_{p-1,p} \circ {\Lambda}_{p-1,p}= \vec \omega_{p-1,p},\quad \vec \omega_{p-1,p} =-2{\Lambda}_{p-1,p}\circ \widetilde{\Lambda}_{p-1,p}^{{}^\centerdot},\quad 
{\Lambda}_{p-1,p}^{{}^\centerdot}=\frac 12\, \vec \omega_{p-1,p} \circ {\Lambda}_{p-1,p}
\label{13cc}
\end{equation}
where $\widetilde{\Lambda}_{p-1,p}=\{{\lambda_0}, -\vec \lambda_{p-1,p}\}$ is {\em conjugation} of $\Lambda_{p-1,p}$.

Let us denote ${\rm col}\{\omega _1, \omega_2, \omega_ 3\}\stackrel{\mathrm{def}}{=}\omega _{p-1,p}^p$, ${\rm col}\{\lambda _1, \lambda _2, \lambda _ 3\}\stackrel{\mathrm{def}}{=}\lambda _{p-1,p}^p$ and ${\rm col}\{\lambda _0, \lambda _{p-1,p}^p\}\stackrel{\mathrm{def}}{=}\Lambda _{p-1,p}^p$
 then 
the orthogonal matrix $C_{p-1,p}$ corresponding to a rotation by the unit quaternion $\Lambda_{p-1,p}$ is given in the following form \cite{Yakovenko}\vspace{3pt}
\begin{equation}C_{p-1,p}(\Lambda_{p-1,p}^p)=
	\begin{bmatrix}
\lambda_0^2+\lambda_1^2-\lambda_2^2-\lambda_3^2&2\lambda_1\lambda_2-2\lambda_0\lambda_3  &2\lambda_1\lambda_3+2\lambda_0\lambda_2  \\
	2\lambda_1\lambda_2+2\lambda_0\lambda_3 &\lambda_0^2-\lambda_1^2+\lambda_2^2-\lambda_3^2&2\lambda_2\lambda_3-2\lambda_0\lambda_1  \\
	2\lambda_1\lambda_3-2\lambda_0\lambda_2  &2\lambda_2\lambda_3+2\lambda_0\lambda_1  &\lambda_0^2-\lambda_1^2-\lambda_2^2+\lambda_3^2\\
	\end{bmatrix}\vspace{7pt}
	\label{RHm}\end{equation}
The quadruple $\Lambda_{p-1,p}^p$ is known as that of Euler--Rodrigues parameters.

From (\ref{13cc}) follows \cite{Yakovenko}
\begin{equation}
	\Lambda_{p-1,p}^{p\centerdot} =\frac 12\, 
	\begin{bmatrix}
	0 & -\omega _1 & -\omega _2 & -\omega _3 \\ 
	\omega _1 & 0 & \omega _3 & -\omega _2 \\ 
	\omega _2 & -\omega _3 & 0 & \omega _1 \\ 
	\omega _3 & \omega _2 & -\omega _1 & 0
	\end{bmatrix} \Lambda_{p-1,p}^p
	\label{RH}\vspace{7pt}
\end{equation}
 Hence equation (\ref{NewtonE}) must be considered along with relations (\ref{13cc})--(\ref{RH}).

From (\ref{RHm}) follows also that there is the matrix $D_{p-1,p}=D_{p-1,p}(\Lambda_{p-1,p}^p)$ such that relation $\omega _{p-1,p}^p=D_{p-1,p}\Lambda_{{p-1,p}}^{p\centerdot}$ is true.\medskip

{\bf 2.2.5. Lagrange equation of II kind}. Let $\lambda _{p-1,p}$ be a triple of Euler angles or Fedorov vector--parameter. Introduce the following notions.
\begin{Definition} \begin{enumerate}\vspace{-7pt}
\item The vectors $q_{p-1,p}={\rm col}\{{\it d}_{p-1,p}^{\it p},\lambda _{p-1,p}\}$ and ${\it q}_{p-1,p}^{\centerdot }={\rm col}\{{\it d}_{p-1,p}^{{\it p}\centerdot },\lambda^{\centerdot }_{p-1,p}\} 
$ are called {\em canonical generalized coordinates} and {\em velocities} of the frame 
${\cal E}_p$ in the motion w.r.t. the frame ${\cal E}_{p-1}$;\vspace{-7pt}
\item the relation \vspace{-21pt}\end{enumerate}
\begin{equation}
V_{p-1,p}^{tw,p}=M_{p-1,p} q_{p-1,p}^{\centerdot },\quad M_{p-1,p}={\rm diag}\{ I,D_{p-1,p}\}
\label{ 1011}
\end{equation}
\hspace{1cm} is called {\em equation of kinematics}\index{Equation!of kinematics} of 
${\cal E}_p$--frame w.r.t. ${\cal E}_{p-1}$.\vspace{-7pt}
\end{Definition}

From relations (\ref{NewtonE}) and (\ref{ 7g}) follows the {\em Lagrange equation}\index{Equation!of Lagrange} of II kind 
\begin{equation} 
{\mathcal A}(q)q^{\centerdot \centerdot }+{\mathcal B}(q,q^{\centerdot })
q^{\centerdot }={\mathcal F}
\label{ 23}
\end{equation}
where ${\mathcal A}(q)=L^TM^TA LM$, ${\mathcal B}(q,q^{\centerdot })=
L^TM^T[A L M^{\hspace{.02cm }{}^\centerdot} +(A L^{\hspace{.02cm }{}^\centerdot} +B)M]$, ${\mathcal F}=L^TM^TF_a$, $M={\rm diag}\{M_{p-1,p}\}$, $q={\rm col}\{q_{p-1,p}\}$.

In the many cases there are constraints on motion of multibody systems, and the matrix $N$ exists such that the matrix $N^TN$ is non--degenerate and we may introduce the generalized coordinate $\tilde q=Nq \in {\bf R}_{m}$ where the natural number $m$ is not more $6k$ \cite{Vel}. 
 Then from relation (\ref{ 23}) follows 
\[\widetilde {\cal A} \tilde q^{\centerdot \centerdot}+\widetilde{\cal B}\tilde q^{\centerdot}= \widetilde {\cal F}
\]
where $\widetilde{\cal A}$, $\widetilde {\cal B}$ and $\widetilde{\cal F}$ are known matrices and column.

As to the quadruple $\Lambda_{p-1,p}^p$, we may replace $\lambda_{p-1,p}$ with $\Lambda_{p-1,p}^p$ in the above definition and equation (\ref{ 23}). 
It is clear that 
 the corresponding matrix ${\mathcal A}$ proves to be singular. Under some assumption this equation is equivalent to a system of differential equations in Cauchy form and algebraic ones. The algebraic equations can be treated as constraints on the multibody system motion. It means that we may introduce `new' coordinates, {\em e.g.}, Euler angles or Fedorov vector--parameter, in order to obtain the Lagrange equation with a non--singular symmetric matrix ${\mathcal A}$.\medskip

{\bf 2.3. A continuum} 
\vspace{3pt}

{\bf 2.3.1. Notion of continuum}.  We shall assume that the sets ${\Lambda_t}\subset \tilde{\Lambda}_t$ are bounded and closed, all their points are continuous, and their surfaces ${\partial{\Lambda_t}}$ are Lyapunov's simple closed surfaces \cite{Zhilin}. 
Let $\mu_2$ be the restriction of $\mu$ on the surface ${\partial{\Lambda_t}}$, $\vec n_x$ be the normal to this surface. 

 Due to \cite{K} constraints being in a small vicinity of $ x\in {\Lambda_t}$ cause  {\em stress}.  Define the internal constraint action as follows \cite{Truesdell,Pobedria}
\[
	{\mathcal F}_{int}({\Lambda} _t)=
	\int \hspace{-.05cm} \chi _{\hspace {-.05cm}_{\partial{\Lambda_t}}}l^{\hskip .02cm{\mathcal T}_x
	 n_x}\mu_2(dx)\vspace{-3pt}
\]
where ${\mathcal T}_x$ is called {\em stress tensor}.

Due to Gauss--Ostrogradsky (divergence) theorem, we have \cite{Pobedria}\vspace{-3pt}
\[{\mathcal F}_{int}({\Lambda} _t)=
		\int \hspace{-.05cm} \chi _{\hspace {-.05cm}_{{\Lambda_t}}}		{\rm div} \hspace{.05cm}l^{\hspace{.02cm}{\mathcal T}_x}		\mu(dx)\vspace{-3pt}\]
Take a point $y(t)$ in a small vicinity of $x(t)\in {\Lambda}_t$ at an instant $t\in {\bf T}$ and define their radius--vectors $\vec r_x(t)$ and $\vec r_y(t)$ (in ${\cal E}_0$) and the vector $\vec h (t)=\vec r_y-\vec r_x (t)$. Then there is the Cauchy--Helmholtz relation 
\cite{Pobedria}
 \[\vec v_y(t)\cong \vec v_x(t)+\frac12[d\vec v_x/d\vec r_x+(d\vec v_x/d\vec r_x)^T]\hskip .07cm\vec h(t)+\frac12[d\vec v_x/d\vec r_x-(d\vec v_x/d\vec r_x)^T]\hskip .07cm\vec h(t)\]
where $\frac12[d\vec v_x/d\vec r_x+(d\vec v_x/d\vec r_x)^T]$ is known as {\em tensor of strain velocities}; $\frac12[d\vec v_x/d\vec r_x-(d\vec v_x/d\vec r_x)^T]$ is known as {\em spin--tensor} at the point $x\in \tilde{\Lambda}_t$ at the instant $t$.

Define the tensor ${\mathcal S}_x(t)$ as the solution of the following equation 
\[
{\mathcal S}_x^{\hspace{.02cm }{}^\centerdot}(t)=\frac12[d\vec v_x/d\vec r_x+(d\vec v_x/d\vec r_x)^T]
\]
with initial data ${\mathcal S}_x={\mathcal I}$, $t=t_0$, ${\mathcal I}$ is the identity (spherical) tensor.

The tensor ${\mathcal S}_x$ is called {\em strain} one \cite{Konoplev1999}. Let us define ${\mathcal U}_x$ as ${\mathcal S}_x $ or ${\mathcal S}_x^{\hspace{.02cm}{}^\centerdot\hskip .05cm}$.
\begin{Definition} The mechanical system $\alpha = \{\sigma _3$, $ \sigma _t$, $\mu$, $\forall t \in {\bf T}$, 
${\Lambda_t}\subset \tilde{\Lambda}_t$, $\forall x\in {\Lambda_t}$, $\rho _x$, $\nu _x$, $ \vec f _x$, $\vec \xi _x$, ${\mathcal T}_x\}$ is called {\em continuous medium} or {\em continuum of Navier--Stocks--Lame class} if the tensor ${{\mathcal T}_x}$ is an {\em isotropic} map of ${\mathcal U}_x$, {\em i.e.}, invariant w.r.t. orthogonal transformations.
\end{Definition}
\medskip

{\bf 2.3.2.  Quasi--linear isotropic matrix--functions}. 

{\bf 2.3.2.1.   3--dimensional case}.
 For any $3\times 3$--matrix $U$ the
aggregate $PUQ$ is an isotropic function of $U$ if the matrices $P$ and $Q$ are
proportional to $I$ with scalar coefficients being invariant w.r.t. rotations. 

 Define the matrices \begin{equation}
E_1=({\rm tr }{\it U})I,\ E_2={\it U},\ E_3={\it U^{T}}\label{ 88}
\end{equation}
where $I$ is the identity matrix. 

Consider the following linear combination 
\begin{equation}
T=r_1{E}_1+{\it r}_2{E}_2+{\it r}_3{E}_3
\label{ 71}
\end{equation}
where $r_i$ are invariant w.r.t. rotations (they can be functions of the time, invariants of $U$ and so on). 
\begin{Theorem} All isotropic quasi--linear $3\times 3$--matrix functions of entries of $U$ are given by relation (\ref{ 71}) {\em \cite{Dubrovin}}.
\end{Theorem}

{\bf 2.3.2.2. 2--dimensional case}. Let $U$ be $2\times 2$ matrix.
It is easy to see that for $2\times 2$
matrices $P$ and $Q$ the aggregate $PUQ$ is an isotropic map of $U$ if $P$ and $Q$ are
of the kind $rI+\widetilde{r}\widetilde{I}$ where  the scalar %
  coefficients $r$ and $\widetilde{r}$ are invariant w.r.t. rotations, $I$ is the identity\vspace{7pt}
  
   $2\times 2$ matrix,
   $\widetilde{I}=\left[
\begin{array}{cc}
0 & -1 \\ 
1 & 0
\end{array}\right] $.\vspace{7pt}

Introduce the following matrices
$E_1=({\rm tr }{\it U})I$, 
$\widetilde{E}_1=({\rm tr }{\widetilde{I}U})\widetilde{I}$,
$E_2={U}$,
$\widetilde{E}_2=\widetilde{I}{U}$, 
$E_3={U}^T$, 
$\widetilde{E}_3={U}^T\widetilde{I}$, 
${E}_4=\widetilde{I}{U}^T$, 
${E}_5={U}\widetilde{I}$, 
${E}_6=\widetilde{I}{U}\widetilde{I}$, 
and 
${E}_7=\widetilde{I}{U^T}\widetilde{I}$. It is easy to see that there are 6 linearly independent matrices, {\em e.g.}, $E_1$, $\widetilde{E}_1$, $E_2$, $\widetilde{E}_2$, $E_3$, and $\widetilde{E}_3$.

Thus there is the set of isotropic quasi--linear $2\times 2$--matrix functions of entries of $2\times 2$--matrix $U$ (invariant w.r.t. ${\mathcal SO}({\bf R}, 2)$)
\begin{equation}
	T =r_1\hskip .05cm({\rm tr }{\it U})\hskip .05cm{I}+\widetilde{r}_1\hskip .05cm({\rm tr }{ \widetilde{I}U})\widetilde{I}+{r}_2{U}+{r}_3{ U}^{T}+\widetilde{r}_2\widetilde{I}{U}+
	\widetilde{r}_3U^{T}\widetilde{I}\label{ 71z}
\end{equation} 
where ${r}_i$ and $\widetilde{r}_i$ are parameters being invariant w.r.t. orthogonal transformations.
\medskip

{\bf 2.3.3. Symmetry of stress tensor}. 
From equation (\ref{ 6}) follows (in the inertial frame ${\cal E}_0$) 
 \cite{Truesdell,Pobedria}
\vspace{-3pt}
\begin{equation}\rho _x
\frac{d}{dt}\hskip .05cm \vec v_x+\nu _x\vec v_x=\rho_x\vec g_x+\vec \xi _x +{\rm div}\hskip .05cm {\mathcal T}_x, \quad {\mathcal T}_x={\mathcal T}_x^T
\label{ 9}\vspace{-3pt}\end{equation}
We may introduce the following constitutive relations with the help of symmetrizing  relations (\ref{ 71}) and (\ref{ 71z}): in the $3$-dimensional case \[{\mathcal T}_x=r_0 {\mathcal I}+r_1\hskip .05cm({\rm tr }\hskip .05cm{\mathcal U}_x)\hskip .05cm {\mathcal I}+{\it r}_2\hskip .05cm{\mathcal U}_x\] and in the $2$-dimensional  case
 \[{\mathcal T}_x=r_0 {\mathcal I}+r_1\hskip .05cm({\rm tr }\hskip .05cm{\mathcal U}_x)\hskip .05cm{\mathcal I}+{\it r}_2\hskip .05cm{\mathcal U}_x+{r}_3(\widetilde{{\mathcal I}}{\mathcal U}_x-
{\mathcal U}_x\widetilde{{\mathcal I}})\] 
where $r_i$ are {\em rheological
coefficients} (parameters being invariant w.r.t. orthogonal transformations); the tensor $\widetilde{\cal I}$ corresponds to the matrix $\widetilde{I}$.\medskip

{\bf 2.3.4. Correct continua}. 
A continuum is called {\em correct} if the corresponding constitutive relation is invertible \cite{Konoplev1999}. With the help of routine calculations we see the following statements to be true:
\begin{enumerate}{
\vspace{-7pt}
\item 
In $3$--dimensional case let $(3r_1+{\it r}_2){\it r}_2\neq 0$. Then there
exists the inverse map \vspace{-7pt}
\[{\mathcal U}_x=n_0{\mathcal I}+n_1({\rm tr }{\mathcal T}_x){\mathcal I} +{\it n}_2\hspace{0.05cm}{\mathcal T}_x\vspace{-7pt}
\] 
where\vspace{-3pt}
\[
n_0 =\frac{r_0}{3r_1+{\it r}_2},\ n_1=-\frac{r_1}{r_2(3r_1+{\it r}_2)}, \ n_2 =\frac 1{r_2}
\]\vspace{-15pt}
\item 
In $2$--dimensional case let $(2r_1+{\it r}_2)( r_2^2+4r_3^2)\neq 0$. Then there
exists the inverse map \vspace{-3pt}
\[{\mathcal U}_x=n_0 {\mathcal I}+n_1\hskip .05cm({\rm tr }{\mathcal T}_x)\hskip .05cm{\mathcal I}+{\it n}_2{\mathcal T}_x+{n}_3(\widetilde{\mathcal I}{\mathcal T}_x-
{\mathcal T}_x\widetilde{\mathcal I})\vspace{-7pt}
\] where\vspace{-3pt}
\[n_0=\frac{-r_0}{2r_1+r_2},\ 
n_1=\frac{-r_1r_2+2r_3^2}{(2r_1+r_2)
({r_2^2+4r_3^2})},\ n_2=\frac{r_2}{r_2^2+4r_3^2},\ n_3=\frac{-r_3}{r_2^2+4r_3^2}\]}\vspace{-17pt}\end{enumerate}
{\bf 2.3.5. Kinds of continua}. If ${\mathcal U}_x={\mathcal S}_x$ and $r_0=0$ the continuum is called {\em elastic material}, if $%
{\mathcal U}_x={\mathcal S}_x^{\hspace{.02cm }{}^\centerdot}$ and $r_0>0$ (called Pascal pressure) the continuum
is called {\em viscous fluid} \cite{Lur'e}.

The continua given above coincide with the continua
used in continuum mechanics in the following cases \cite{Truesdell,Lur'e}
\begin{quote}{\parskip -.02cm \vspace{-33pt}
\item  \hskip -.8cm  --- \thinspace \thinspace  \thinspace \thinspace {\em
the Pascal pressure $r_0$ is positive and $r_1=r_2=r_3=0$ ({ideal fluid});\vspace{-3pt}
\item  \hskip -.8cm  --- \thinspace \thinspace  \thinspace \thinspace $r_0$ is non--negative and $r_1{\rm tr }{\mathcal I}+{\it r}_2\neq 0\rightarrow 
{\rm tr }\hspace {.05cm} {\mathcal T}_x\neq -{\it r}_0{\rm tr }{\mathcal I}$ (correct continua) (here $\mathcal I$ is used as $2$-- and $3$--dimensional identity tensors, respectively).}\vspace{-3pt}
}\end{quote} 

{\bf 2.4. Systems with inhomogeneous screw measures}

 Show how the systems with inhomogeneous  sliders can be realized in the
conventional mechanics.

{\bf 2.4.1.  Multiphase systems}. 
Equation (\ref{ 7}) is realized for multiphase systems (equations (6.34) and (7.11), given in \cite{Pobedria}, can be written in the form of (\ref{ 7})). Here the stress tensor proves to be non--symmetrical \cite{Pobedria}, and the motion equations are six--dimensional.
\medskip

{\bf 2.4.2. Elements of Eulerian mechanics}.  The mechanical sense of the vectors $\vec p_x$ and $\vec q_x$ in equation (\ref{ 7}) may be clarified in the framework of {\em Eulerian mechanics} \cite{Zhilin} (we are not going to discuss its meaning as a base of mechanics). 

A particular case of bodies is the well--known {\em mass--point} being {the fundamental concept of theoretical mechanics.  It is considered as the unique model of a natural things having infinitesimal sizes, but possessing masses. Is such model a universal one? To answer this question, it is necessary to address to physics. The modern physics draws the following picture of a material objects: it consists of molecules, atoms, protons, neutrons, electrons, neutrinos or from their aggregates which are called clusters. What of these objects leads to the concept of a mass--point of theoretical mechanics? Let us take, for example, an electron. Its sizes are extremely small, it possesses some mass, so as though, it may be modeled as a mass--point. But here that disturbs us. It has appeared that at decoding and interpretation of tracks of nuclear particles, including electrons after their collisions, it is necessary to consider spins of these particles, to be exact, their angular momentums. Angular momentum is connected with rotation of these particles. But by definition a mass--point cannot rotate.
It means that even a such small object as an electron cannot be modeled as a mass--point. Let us take a larger object, for example, a cluster or crystallite of some polycrystalline metal. Certainly, it is possible to model motion of its center of masses as motion of some mass--point having the same mass, as well as the mass of cluster or crystallite. But a cluster or crystallite can rotate round the center of masses. Thus a cluster cannot be modeled as a mass--point, too.
 
That is why  Eulerian mechanics supplies  points of the sets $ {\Lambda_t} $, $t \in {\bf T}$, with translation velocities $\vec v_x$ and angular ones $\vec{\omega}_x\in {\bf V}_3$, as well as densities $ A_x$, $ B_x$ and $ C_x$ of generalized inertia tensors. Then the kinetic energy (\ref{K}) is introduced by its positive defined density $k_x\stackrel{{def}}{=}\frac{1}{2}\langle \vec{v}_x, A_x \vec{v}_x\rangle +\langle \vec{v}_x, B_x \vec{\omega}_x\rangle+
	\frac{1}{2}\langle \vec{\omega}_x, C_x\vec{\omega}_x\rangle$. After that one defines the vectors $\vec p_x\stackrel{{def}}{=}\frac{\partial}{\partial \vec{v}_x} k_x=A_x\vec{v}_x+B_x\vec{\omega}_x$ and  $\vec q_{y}\stackrel{{def}}{=}r_{yx}^\times \vec p_x + \vec q_x$ where 
$\vec q_x{=}\frac{\partial}{\partial \vec{\omega}_x} k_x=B^T_x\vec{v}_x+ C_x\vec{\omega}_x$ is the density of so called {\em dynamical spin}.

With the help of $\vec p_x$ and $\vec q_x$ the slider $l^{ p_x, q_x}$ is introduced, equation (\ref{ 7})  is postulated  \cite{Zhilin}.

Realization of equation (\ref{ 7}) in Euler mechanics is motion equations of point--bodies and their systems, thin rods and so on \cite{Zhilin}. 

\medskip

{\large \bf Conclusion} 





In mechanics there is mainly absent the 
 understanding that motion of bodies and  interaction between them can be described with the help of screws as 
it is considered as conventional that `$\ldots$being very attractive representation of a system of forces and rigid body motions with the help motors and screws, nevertheless it has no essential practical value$\ldots$' \cite{Sommerfeld} and that the screw calculus is not adapted for the description of continuum motion \cite{Dimentberg}.

At the same time screw calculus gives useful, convenient and necessary tools which permit us to postulate the fundamental principle of dynamics in the   differential form (see  \cite{Konoplev1999} and auhtor's paper `On Foundations of Newtonian Mechanics',   arXiv:1012.3633). However this form 
 leaves in a shade many important features of rational mechanics that can be understood only with using the  (stronger) local (primitive) integral form of the conservation (change) law
 for the vector  measuare of motion  (the differential form  is applicable only in the cases where the  divergence  theorem is true).





 
 In order to obtain the integral form, the new notions of homogeneous and inhomogeneous vector and tensor slider--functions and screw measures are used, and  the main mechanics measures, the equation of motion and the concept of mechanical system are introduced. 
 It is shown that 
 mass--points, rigid bodies, continua, multiphase systems,  and point--bodies  are  realizations of mechanical systems of the given axiomatics (see also \cite {Berthelot,Pobedria,Zhilin}). 
 

In this way we solve also the following problem:
 
 `$\ldots$ the dynamics of a continuous system must clearly include as a
limiting case (corresponding to a medium of density everywhere zero except
in one very small region) the mechanics of a single material particle. This
at once shows that it is absolutely necessary that the postulates introduced
for the mechanics of a continuous system should be brought into harmony with
the modifications accepted above in the mechanics of the material particle' \cite{Levi--Civita}. 



 The author would be highly grateful with whoever would bring any element likely to be able to make progress the development, and thus the comprehension, of the paper. Any comments, 
  critiques, or objections are kindly invited to be sent to the author by e--mail.


{

\end{document}